\documentclass[reprint,superscriptaddress,frontmatterverbose,showpacs,showkeys,preprintnumbers,onecolumn,notitlepage,
amsmath,amssymb,aps,pra,floatfix,amsfonts,longbibliography]{revtex4-1}
\usepackage{graphicx,amsmath,amsfonts,amssymb,upgreek,txfonts,color}
\usepackage{xcolor}
\usepackage{bm}
\usepackage{bbm}
\usepackage[pagebackref=false,colorlinks,linkcolor=red,citecolor=blue,urlcolor=red,breaklinks=true]{hyperref}
\usepackage[capitalise]{cleveref}

\colorlet{acolor}{red!60!black}

\usepackage{soul}
\usepackage{bbding}

\newcommand*{\scrpt}[1]{\mathrm{#1}}
\newcommand*{\s}[1]{\ensuremath{_\scrpt{#1}}}
\newcommand*{\up}[1]{\ensuremath{^\scrpt{#1}}}

\newcommand*{\ket}[1]{\left|#1\right\rangle}

\newcommand{\mean}[1]{\left\langle #1 \right\rangle}
\newcommand*{\coherence}{\mathcal{C}}
\newcommand*{\Cq}{\coherence_q}
\newcommand*{\Cm}{\coherence_m}

\begin{document}

\newcommand*{\UPOL}{Department of Optics, Palack{\'y} University, 17.~Listopadu~12, 771~46~Olomouc, Czech~Republic }
\title{Thermally-induced qubit coherence in quantum electromechanics}
\author{N. \surname{Etehadi Abari} }
\email[Corresponding author: ]{najmeh.etehadiabari@upol.cz}
\affiliation{\UPOL}
\author{A. \surname{Rakhubovsky}}
\email{rakhubovsky@optics.upol.cz}
\affiliation{\UPOL}
\author{R. \surname{Filip}}
\email{filip@optics.upol.cz}
\affiliation{\UPOL}
\date{\today}
\begin{abstract}
    Quantum coherence, the ability of a quantum system to be in a superposition of orthogonal quantum states, is a distinct feature of the quantum mechanics, thus marking a deviation from classical physics.
    Coherence finds its applications in quantum sensing and metrology, quantum thermodynamics and computation.
    A particularly interesting is the possibility to observe coherence arising in counter-intuitive way from thermal energy that is without implementation of intricate protocols involving coherent driving sequences.
    In this manuscript, we investigate quantum coherence emerging in a hybrid system composed of a two-level system (qubit) and a thermal quantum harmonic oscillator (a material mechanical oscillator), inspired by recent experimental progress in fabrication of such systems.
    We show that quantum coherence is created in such a composite system solely from the interaction of the parts and persists under relevant damping.
    Implementation of such scheme will demonstrate previously unobserved mechanisms of coherence generation and can be beneficial for hybrid quantum technologies with mechanical oscillators and qubits.
\end{abstract}

\keywords{}

\maketitle

\section{Introduction}\label{sec:1}

Coherence is a fundamental concept in quantum mechanics that is connected to the superposition of quantum states in a basis preferred for a certain application.
Quantum states that possess this non-zero coherent superposition of basis states can provide advantage for science and technology over the incoherent statistical mixtures of the same basis states.
Coherence enhances performance of the quantum protocols in sensing and metrology~\cite{giovannetti_advances_2011,degen_quantum_2017}, quantum thermodynamics~\cite{misra_energy_2016,santos_role_2019}, and quantum information processing~\cite{galindo_information_2002,ladd_quantum_2010,matera_coherent_2016}.
Quantum coherence has been shown to play a role in biological processes as well~\cite{lloyd_quantum_2011,ishizaki_quantum_2012}.
In order to quantify the coherence, a few resource theories have been put forward~\cite{winter_operational_2016,streltsov_colloquium_2017,bischof_resource_2019,smith_quantum_2022}.
Interplay between coherence and other quantum resources such as entanglement, discord and steering has been investigated in~\cite{yao_quantum_2015,hu_quantum_2018}.
On the other hand, it remains unexplored how quantum coherence emerges during quantum dynamics from incoherent thermal states.

Generally, quantum coherence of an open system emerges in presence of an external strong coherent drive.
Recently, it has been shown~\cite{guarnieri_steadystate_2018} that quantum coherence can emerge in a steady state of a system that only interacts with its environment given certain properties of this interaction.
Subsequent studies proposed similar system-environment phenomena~\cite{guarnieri_nonequilibrium_2020,reppert_equilibrium_2020,roman-ancheyta_enhanced_2021,cresser_weak_2021,slobodeniuk_extraction_2022,cerisola_quantumclassical_2022}.
In parallel, an experimental proposal in double-quantum-dot solid-state systems was analyzed~\cite{purkayastha_tunable_2020a}.
However, even proof-of-principle experimental tests of such phenomena are still missing due to the challenging engineering of composite interactions.

In our work we investigate coherence emerging in a hybrid electromechanical system similar to the one studied in~\cite{ma_nonclassical_2021}.
We show that coherence in each subsystem can emerge solely from coherent interaction between the constituents that start from fully incoherent states.
We analyze such thermal rise of quantum coherence in quantum electromechanics and propose an experiment to observe the principle mechanism.
Moreover, we describe a regime where thermal mechanical oscillator monotonously stimulates qubit coherence, even if the phonon number is much larger than unity.
Hybrid systems such as this combine benefits of the constituents allowing the positive synergy to open new perspectives in science and technology.
Electromechanical systems, by combining advantages of superconducting devices and high-Q mechanical oscillators, allow preparation of exotic states of macroscopic mechanical oscillators~\cite{oconnell_quantum_2010,ma_nonclassical_2021}, and transduce quantum information between microwave and optical domains~\cite{brubaker_optomechanical_2021,sahu_quantumenabled_2022}.
Such transduction not only allows an effective long-range communication between superconducting devices but also their effective readout by optical means~\cite{delaney_nondestructive_2021}.

\section{Results}\label{sec:2}
\subsection{Model of the qubit-mechanical system}\label{sub1:qub-mech-model}

\begin{figure}[hbt!]
    \centering
    \includegraphics[height = 7cm]{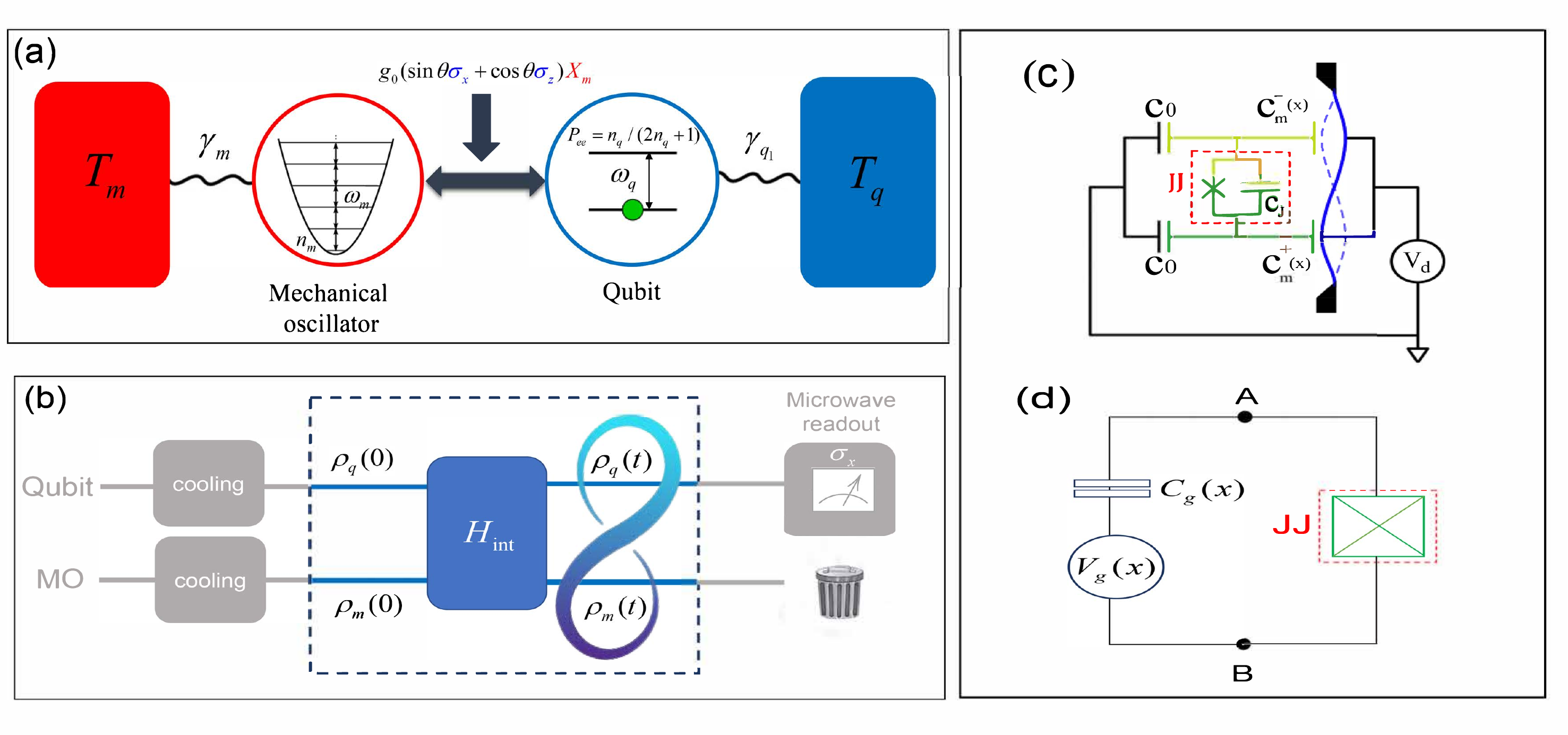}
    \caption{(a) Schematic diagram of the physical system. A single-mode mechanical harmonic oscillator of frequency $\omega_m$ is coupled to a qubit (frequency $\omega_q$) via a general coupling rate $g_0$.
        (b) A sketch of the interaction protocol between the qubit and the mechanical mode.
        Before the interaction, the qubit and the mechanical oscillator are prepared in the incoherent states, respectively, $\rho_q(0) $ and $\rho_m (0)$, either by cooling, or by equilibration with the corresponding bath.
        The quantum coherence is evaluated after the interaction has finished, and can be probed by microwave readout.
        (c) An experimental illustration of the model consisting of a charge qubit (CPB) coupled to the mechanically compliant capacitors in an electromechanical system~\cite{ma_nonclassical_2021}. The red-dashed rectangle area indicates the Josephson Junction (JJ) represented by a nonlinear inductor and a Josephson capacitor $C_J$. The suspended superconducting islands of the CPB which connect the charge qubit to the superconducting reservoir (other parts of the circuit) are displayed in light and dark green colors. The motion of the mechanical oscillator (blue electrodes) can modify the separation between the two capacitors $C_m^{\pm}(x)$ which are modulated with the opposite phase by the anti-symmetric motion of the mechanical oscillator (MO). $V_{\mathrm{dc}}$ characterizes the DC-voltage applied to the MO. The gate-charge (offset charge) $n_g = C_g(x) V_g(x) /2 $ applying on the CPB, can be defined by the equivalent capacitor $C_g (x)$ and voltage $V_g(x)$ of the circuit which are now position-dependent (see Appendix~\ref{A1}). The modulation of the offset charge via the mechanical motion induces a coupling between the mechanical motion and qubit-electrostatic energy. (d) The equivalent circuit of the experimental model.
    }
    \label{fig1}
\end{figure}

In this manuscript, we demonstrate a possibility to generate coherence in a coupled system of nanomechanical oscillator and a two-level system (a qubit) from a fully incoherent state.
A schematic depiction of the scheme is in~\cref{fig1}~(a).
First, we introduce a theoretical description of the system and the figures of merit.
The electromechanical systems of interest, akin to investigated in~\cite{ma_nonclassical_2021}, can be described by the Hamiltonian ($ \hbar \equiv 1$)
\begin{equation}
    \label{eq:hamiltonian_first}
    H = \frac{ \omega_q }{2 } \sigma_z + \frac{ \omega_m }{2 } ({X}_m^2 + {P}_m^2) + \sqrt{2}  g_0 ( \sin\theta  \sigma_x + \cos \theta \sigma_z ) {X}_m.
\end{equation}
Here the first two terms describe the free dynamics of the qubit (with Pauli matrices $ \sigma_i$ and transition frequency $\omega_q$) and the nanomechanical oscillator (with eigenfrequency $\omega_m$ and the dimensionless position and momentum quadratures, respectively, ${X}_m$ and ${P}_m$ normalized such that $[{X}_m, {P}_m] = i$).
For convenience, we also define the detuning $\Delta = \omega_q - \omega_m$.
The third term in~(\ref{eq:hamiltonian_first}) describes the interaction between the qubit and mechanics
required to achieve emerging quantum coherence~\cite{guarnieri_steadystate_2018}.
We focus on proof-of-principle demonstration of the interaction mechanism using only one dominant mode at the frequency $\omega_m$ coupled to an external bath.
From a thermal occupation of the qubit, this composite interaction can generate a coherent displacement of the oscillator, continuously generating quantum coherence in the qubit.
In an experiment the hybrid interaction can be realized via capacitive, magnetic flux or electromotive coupling methods~\cite{xiang_hybrid_2013}.
The coupling can be tuned in magnitude by changing the rate $g_0$ or adjusted by manipulating the value of $\theta$.
This can be advantageously reached by utilizing the suitable lumped elements in the superconducting circuit~\cite{ma_nonclassical_2021,xiang_hybrid_2013,girvin_circuit_2009} since in our model $\theta$ depends on the charging and Josephson energies while $g_0$ can be controlled through DC voltage bias and capacitors of the circuit as well as charging energy (see Fig.~\ref{fig1}(c,d) and Appendix~\ref{A1} for more details).

To investigate emerging coherence in such system, we assume that both mechanics and qubit are prepared initially in thermal states, states that lack coherence in the natural basis of Fock states.
The initial state of the compound system therefore reads
\begin{equation}\label{rho_i}
   {\rho}(0) =
    {\rho}_{\mathrm{qubit}}(0) \otimes {\rho}_{m}(0) =
    \Big(P_{ee} \vert e \rangle \langle e \vert  + ( 1 - P_{ee}) \vert g \rangle \langle g \vert \Big) \otimes
    \sum_{k = 0 }^{\infty} \frac{n_m^k}{( 1 + n_m )^{k + 1}} \vert k \rangle \langle k \vert ,
\end{equation}
where $\ket{g}[\ket e]$ is the ground [excited] state of the qubit, $\ket k$ is a Fock state of the mechanical oscillator, $P_{ee} = n_q / ( 2 n_q + 1 )$.
The mean occupation number of mechanics $n_m$ and the occupation parameter $n_q$ of the qubit obey Bose-Einstein statistics: $n_i = [ \exp ( \hbar \omega_i / k\s B T_i) -1]^{-1}$  for $i = q,m$, with $k\s{B}$ being the Boltzmann constant and $T_i$ the temperature of the corresponding subsystem.

The dynamics generated by the Hamiltonian~(\ref{eq:hamiltonian_first}) is capable of driving the initially incoherent state~(\ref{rho_i}) into a state in which both mechanics and the qubit possess quantum coherence.
From a plethora of available measures of coherence (see Ref.~\cite{streltsov_colloquium_2017} for a review), we choose the $l_1$-norm-based measure~\cite{baumgratz_quantifying_2014} to quantify the qubit coherence.
This measure has the meaning of mean displacement in $xy-$plane and can be computed for the qubit as
\begin{equation}
    \label{eq:coherence_def}
    \Cq = \sqrt{ \mean{ \sigma_x \vphantom{\sigma_y} }^2 + \mean{ \sigma_y }^2}.
\end{equation}
Throughout the manuscript we will compare the qubit's coherence with the mean coherent displacement of the oscillator $\Cm = \sqrt{ \mean{{X}_m}^2 + \mean{{P}_m}^2}$.
Note that in general the $l_1$-norm such as displacement is not a proper coherence monotone for an oscillator (a system with infinite-dimensional Hilbert space) as it can diverge on states with finite mean energy~\cite{zhang_quantifying_2016}.
Nevertheless, the mean coherent displacement is an illustrative quantity that can provide a quantum advantage in e.g. metrology.
The mean values in~\cref{eq:coherence_def} are computed over the evolved quantum state $\rho(t)$.
In the case of unitary dynamics, $\rho(t) = e^{ - i H t  } \rho(0) e^{ i H t}$.
In a realistic case where both systems are subject to decoherence caused by interaction with the corresponding environment, one has to use more complicated tools, such as solving master equation (see~\cref{sec:methods} for elaboration).
\begin{figure}[t!]
  \centering
  \includegraphics[width=5.5cm]{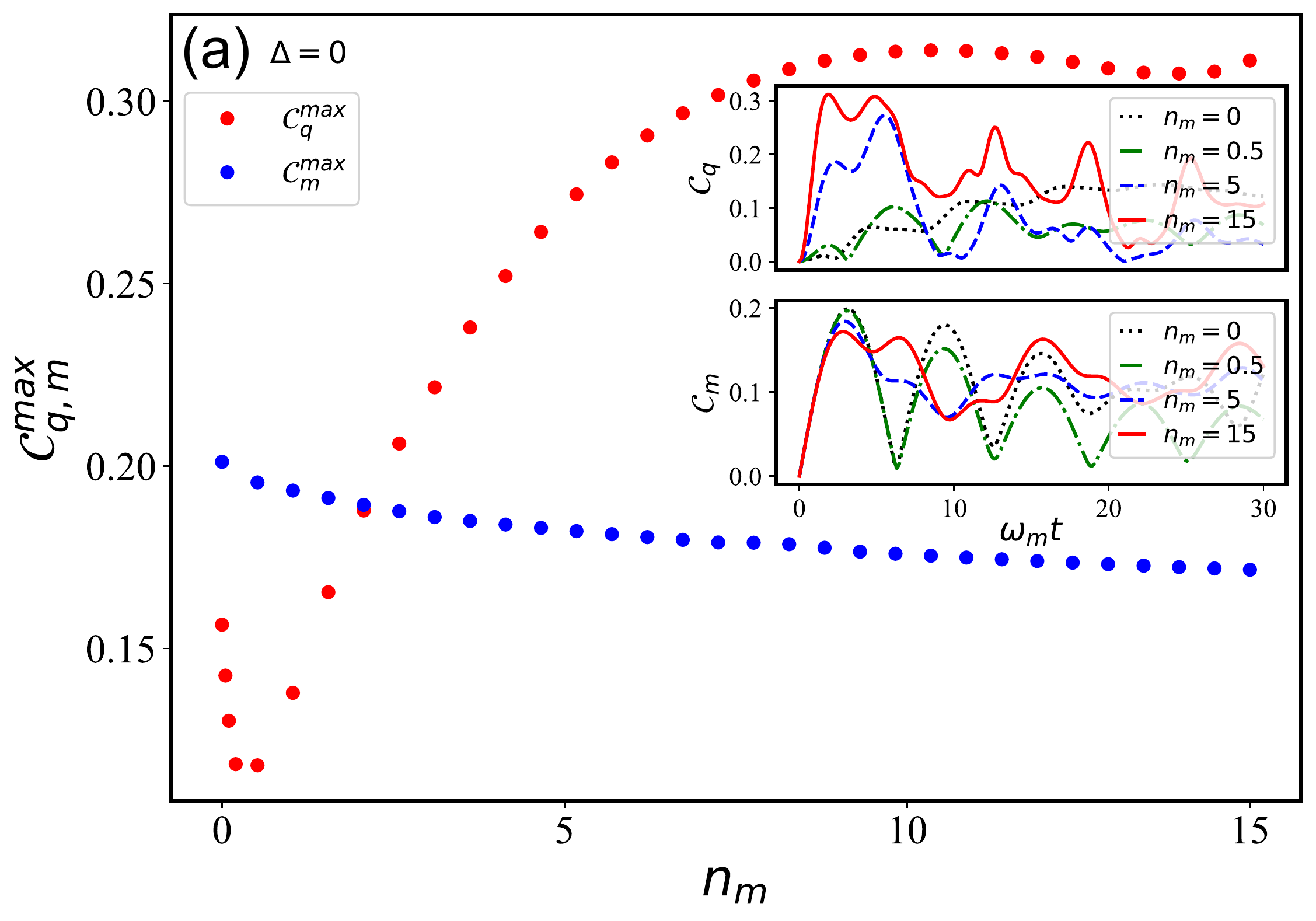}
  \hspace*{0.008cm}
  \includegraphics[width=5.5cm]{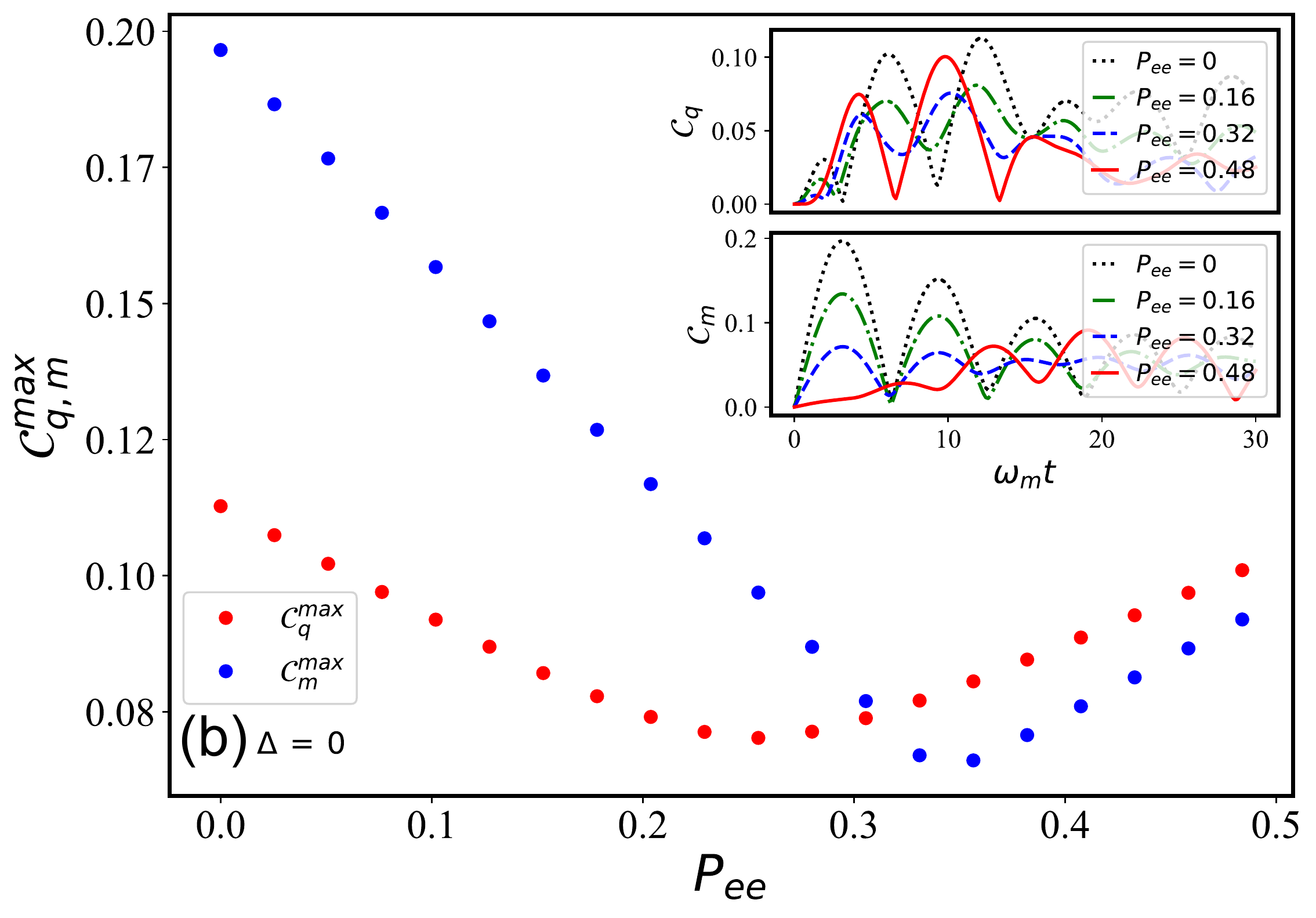}
  \hspace*{0.008cm}
  \includegraphics[width=5.5cm]{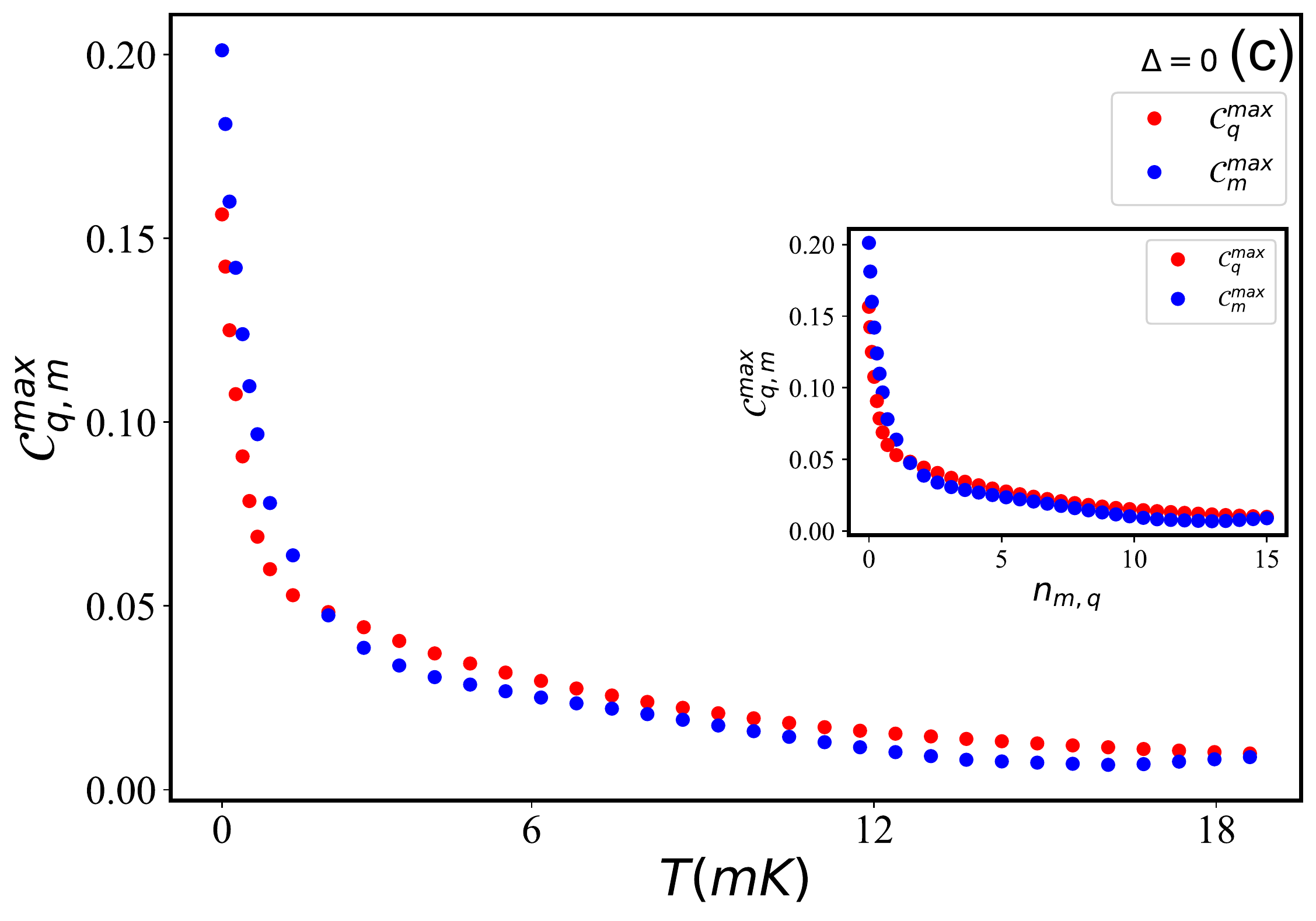}
  \\
  \includegraphics[width=5.5cm]{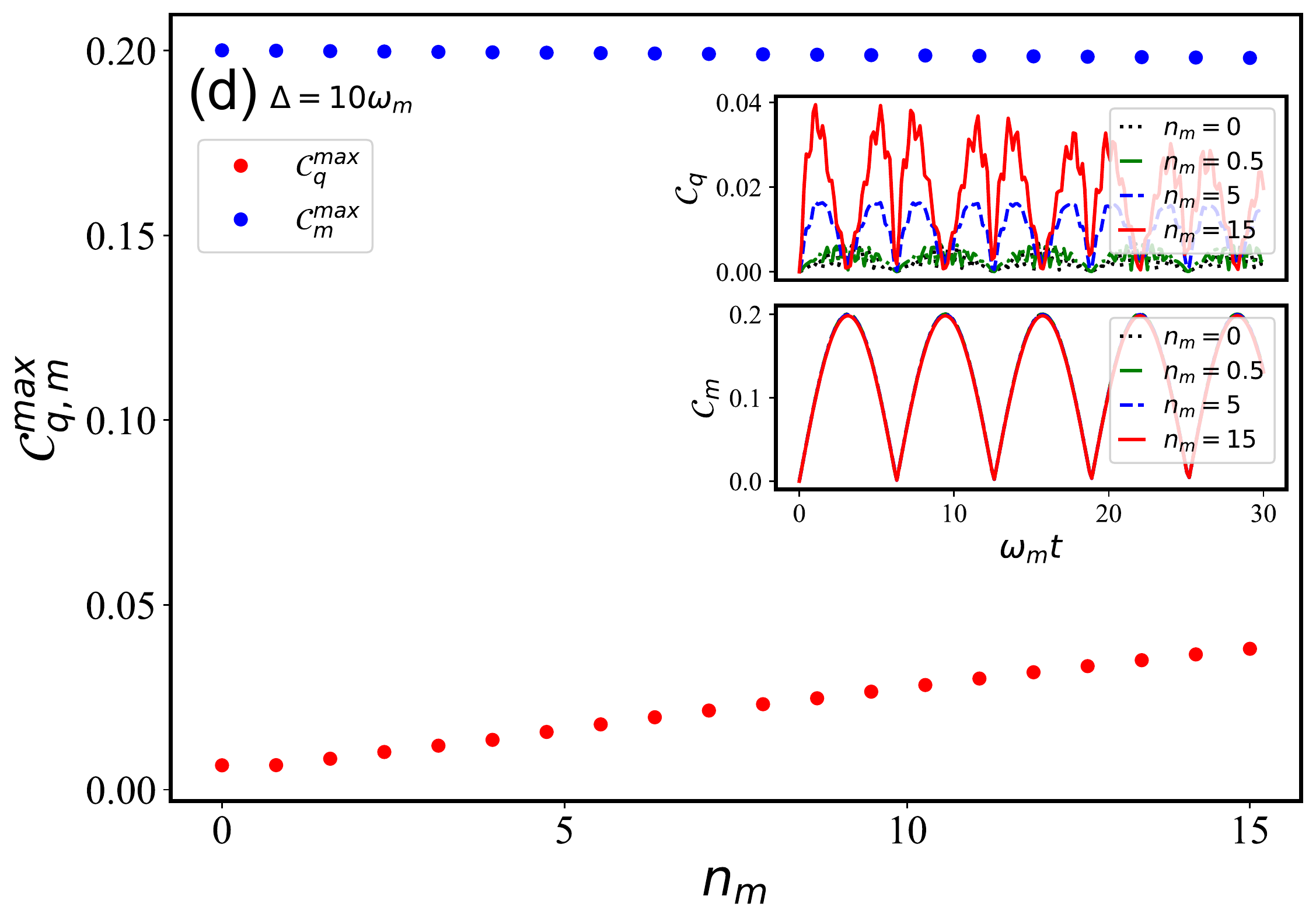}
  \hspace*{0.008cm}
  \includegraphics[width=5.5cm]{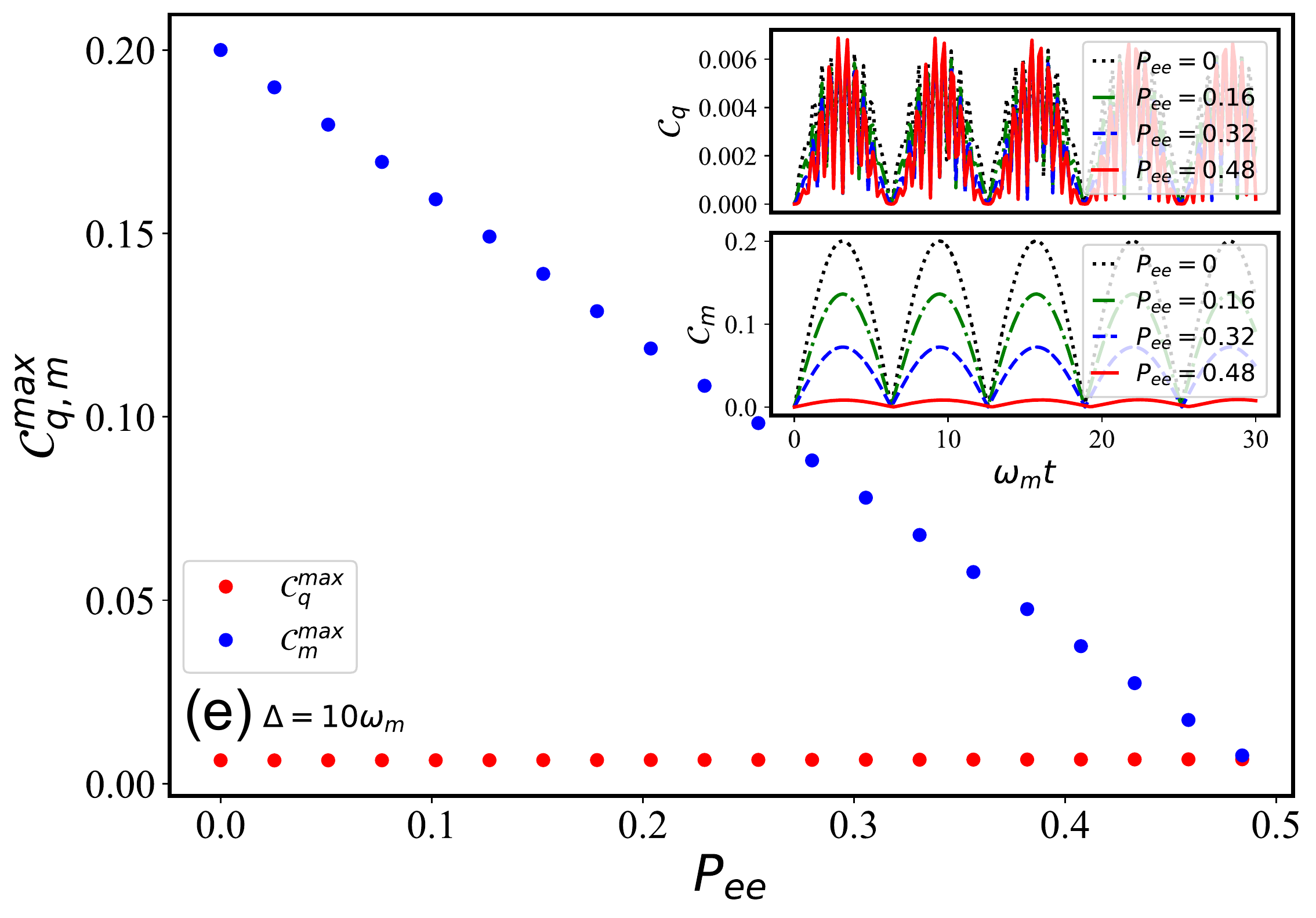}
  \hspace*{0.008cm}
  \includegraphics[width=5.5cm]{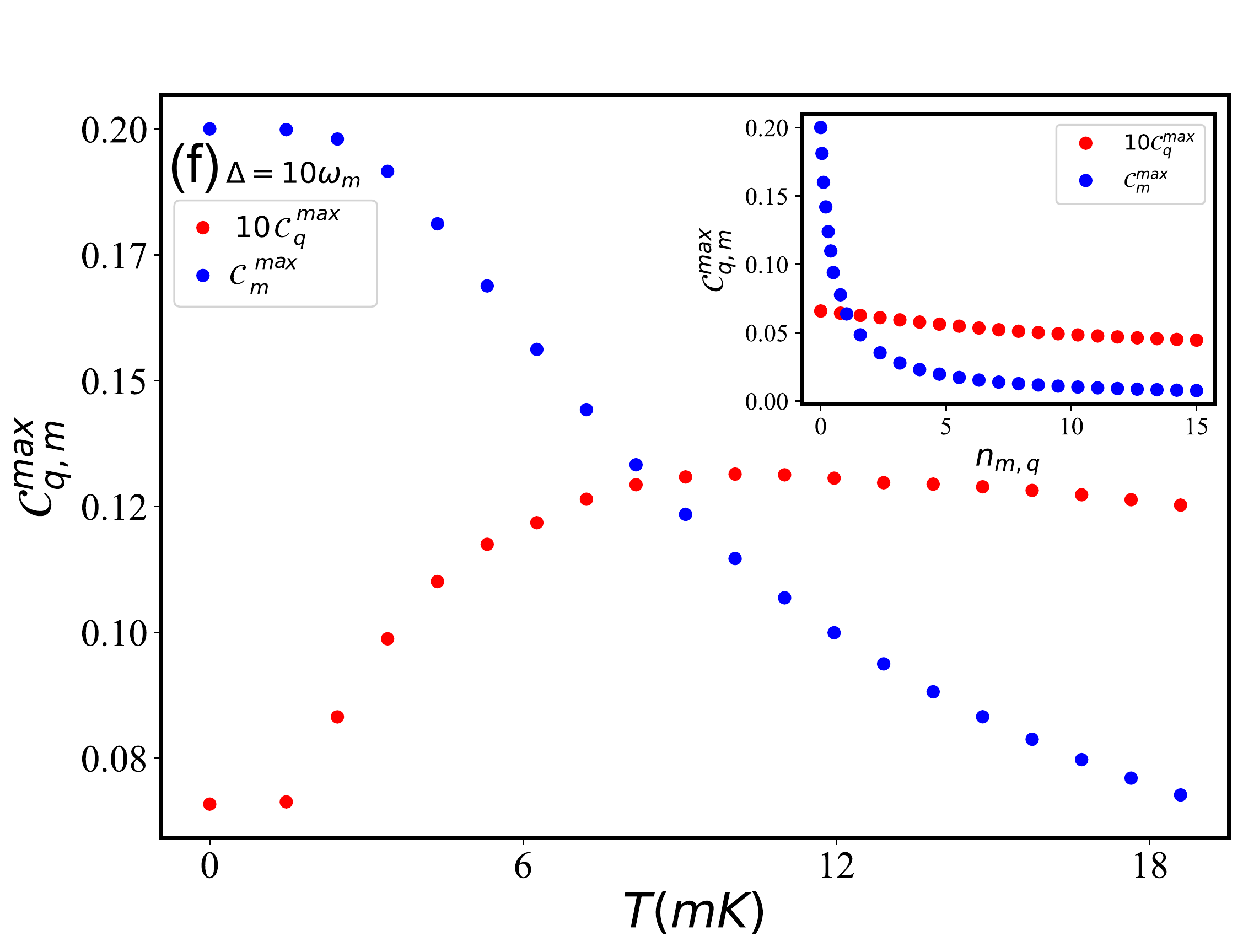}
  \caption{Maximally attainable coherence $\Cq$ and mechanical displacement $\Cm$ as a function of the initial occupation of mechanics (a,d) or qubit (b,e) given a constant initial occupation of the other subsystem.  In (a,d) the qubit is initially in the ground state $P_{ee} = 0$. In (b,e) the mechanics has initial occupation $n_m = 0.5$. Insets show the evolution of $\mathcal{C}_q$ and $\mathcal{C}_m$ as functions of time for different occupation.
    Note that $\Cq$ and $\Cm$ assume their corresponding maximal values at different instants of time.
    (c,f) Optimum values of $\Cq, \Cm$ as a function of the initial temperature assuming equal temperature baths for both subsystems.
    The inset plots of panels (c,f) show how $\Cq\up{max}, \Cm\up{max}$ change as a function of equal initial occupation (in this case, the initial temperatures differ in panel (f)).
    In each panel, weak coupling regime $g_0 = 0.1 \omega_m$ is assumed.  The panels (a,b,c) correspond to the resonance between the qubit and MO ($\omega_m = \omega_q$), in (d,e,f) $\omega_q - \omega_m = \Delta = 10\omega_m$.
  }
  \label{fig2}
\end{figure}
\subsection{Quantum coherence generated by pulsed noiseless dynamics} % <<<
\label{sec:coherence_generated_by_unitary_dynamics}

The simple model of the Hamiltonian~\eqref{eq:hamiltonian_first} captures a rich dynamics whose exact type depends on the interplay between the eigenfrequencies of individual subsystems $\omega_{m,q}$ and the coupling defined by its magnitude $g_0$ and phase $\theta$.
Moreover, generation of coherence in this system is determined by the initial state before the interaction starts.
In this subsection, we show that, counterintuitively, increasing temperature of the initial quantum state can be beneficial for generation of coherence in the qubit.
To estimate the limits of attainable values of coherence, we start with the noiseless case when the two subsystems, the MO and the qubit are decoupled from their environments and only couple to each other.

In order to see the effect of the initial thermal occupation on the coherence generation, we simulate the dynamics of the system driven by only the Hamiltonian~\eqref{eq:hamiltonian_first} and ignore the coupling to the environment.
In this case, the quantum state of the bipartite system after the interaction can be obtained straightforwardly by applying unitary transformation to the initial product state~\eqref{rho_i}.
The estimates of the coherence emerging from the unitary qubit-mechanical interaction are shown at~\cref{fig2}.
The numerical study assumes weak coupling regime $g_0 = 0.1 \omega_m \ll \omega_m + \omega_q$, and $\theta = \pi /4$, equal coupling of mechanical displacement to both $\sigma_x$ and $\sigma_z$, in order to gain the optimum values of the coherence parameters (the dependence of the coherence parameters to $\theta$, i.e., coupling rates $g_x$, $g_z$, as well as the absolute value of the qubit-mechanical coupling $g_0$ and the detuning $\Delta$ is discussed in more details in Sec.~\ref{sub3:Eff-coupl-detun}).

As seen in Fig.~\ref{fig2}(a,d), having a hotter initial mechanical state has a positive effect on qubit coherence such that by increasing the mechanical temperature or equivalently increasing thermal occupation $n_m$, we reach higher maximum values for $\mathcal{C}_q\up{max}$.
The duration of time it takes to reach the maximal value $\Cq\up{max}$ is also reduced with increasing initial occupation $n_m$, which is illustrated by the inset plots.
This phenomenon contrasts with a steady-state qubit coherence induced by a multimode bosonic bath~\cite{guarnieri_steadystate_2018,purkayastha_tunable_2020a,slobodeniuk_extraction_2022}, where the maximum of coherence appears for vanishing temperature.
Interestingly, the increase in coherence is accompanied by only a moderate coherent displacement decrease in the oscillator.
The opposite happens when the qubit's initial temperature is increased at $\Delta = 0$ when we fix the value of $n_m = 0.5$. As seen in Fig.~\ref{fig2}(b) the maximum accessible amounts of $\mathcal{C}_q$ can be reached when $P_{ee} = 0$, i.e., $n_q = 0$. Elevated initial occupations of the qubit do not significantly alter the qubit coherence $\mathcal{C}_q$ in the dispersive regime $\Delta = 10 \omega_m$.
Therefore, it is advantageous to keep the qubit initially in the ground state and increase the oscillator's initial temperature to observe emerging quantum coherence, more significant than the steady-state coherence~\cite{guarnieri_steadystate_2018,purkayastha_tunable_2020a,slobodeniuk_extraction_2022}.

The optimum values of $\mathcal{C}_m$ show a slow reduction as a function $n_m$ for $\Delta =0$, while in the dispersive regime the maximum values of $\mathcal{C}_m\up{max}$ do not change considerably with respect to $n_m$ (compare Fig.~\ref{fig2}(a,d), blue dots, and inset plots for $\mathcal{C}_m$). In addition, by increasing the detuning and moving from the resonance case to the off-resonance one,  the decrease rate of $\mathcal{C}_m\up{max}$ becomes faster when the qubit temperature rises (compare Fig.~\ref{fig2}(b,e), blue dots, and inset plots for $\mathcal{C}_m$).

Finally, for the case in which the initial temperatures of the qubit and the MO are equal ($T_m = T_q = T $), the maximum attainable amounts of $\Cq$ and $\Cm$ are shown in panels (c) and (f) of Fig.~\ref{fig2}, for resonance and off-resonance cases, respectively.
Inset plots of Fig.~\ref{fig2}(c,f) also demonstrate the optimum values of coherence parameters as a function of the initial occupation assumed equal for both subsystems ($n_q = n_m = n_{m,q}$).
As is seen, by increasing the thermal occupation numbers of two subsystems at the same time, i.e., increasing $n_{m,q}$, $\Cq\up{max}$, and $\Cm\up{max}$ decrease (the reduction rate of the $\Cq\up{max}$ as a function of $n_{m,q}$ is not significant in dispersive regime $\Delta = 10 \omega_m$).
For the case of resonance, the results of the inset plots are the same as the main plot (c), as $\omega_m = \omega_q$ and $T_m = T_q = T$ give us the identical occupations $n_m = n_q = n_{m,q}$.
However, at $\Delta = 10 \omega_m$, the main plot of Fig.~\ref{fig2}(f) for $\Cq\up{max}$ shows a small  increase  as the temperature of the baths rises simultaneously.
Therefore, we can conclude that as long as $n_q < n_m$, by raising the temperature, it is possible to observe an increase of the value of the qubit coherence parameter.

In addition, by comparing the first row and the second row of Fig.~\ref{fig2}, we realize that by increasing the detuning, the energy exchange between the mechanical mode and qubit through the coupling channel $g_x = g_0 \sin \theta$ reduces, which causes the reduction in maximum accessible amount of qubit coherence since $\mathcal{C}_q$ depends on both $g_x$ and $g_z = g_0 \cos \theta$ (see Sec.~\ref{sub3:Eff-coupl-detun} for further details).
On the other hand, as $\mathcal{C}_m$ is only influenced by coupling rate $g_z$, increasing the detuning does not affect the maximum reachable amount of $\mathcal{C}_m\up{max}$.
\begin{figure}[t!]
\centering
\includegraphics[width=5.9cm]{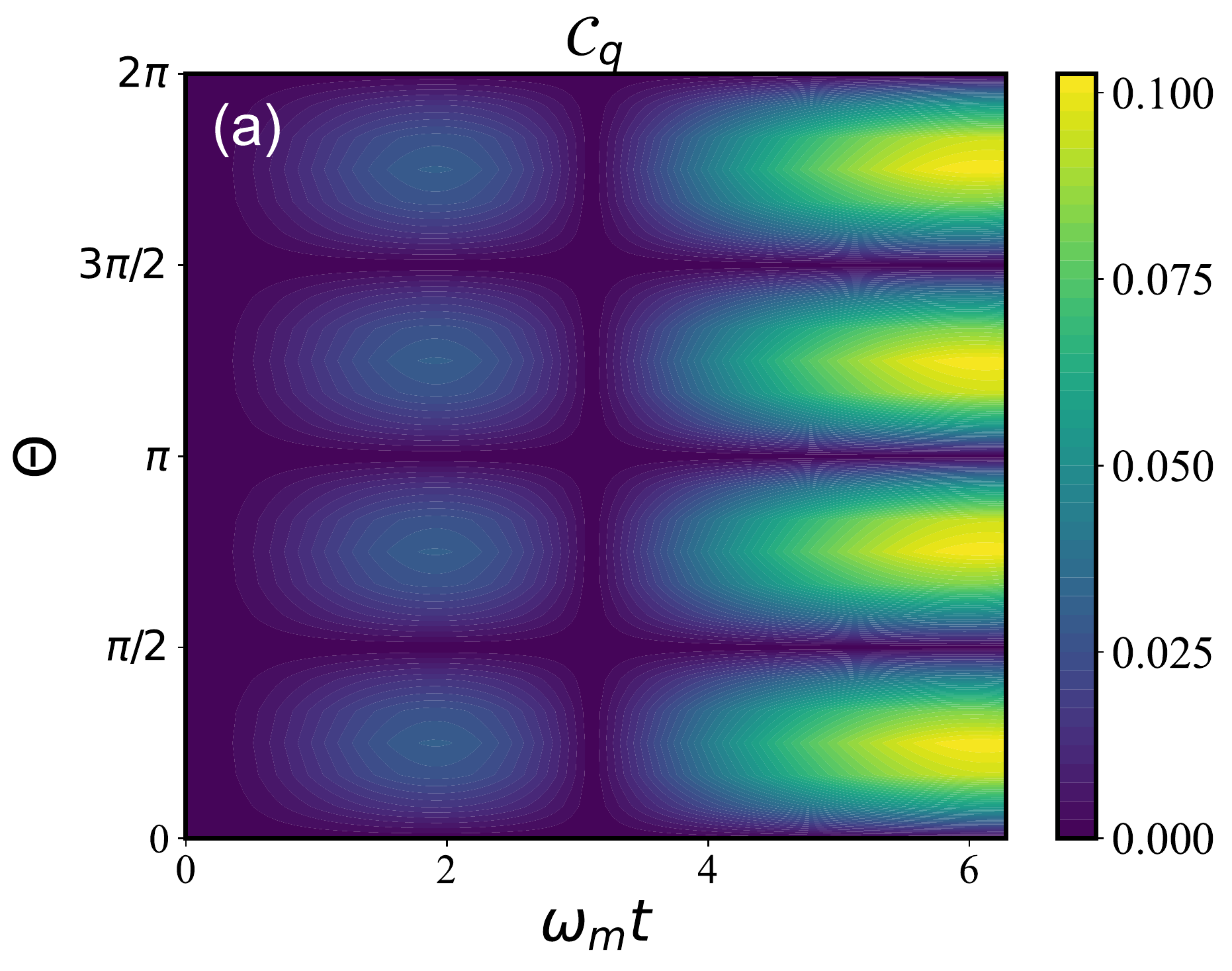}
\hspace*{0.008cm}
\includegraphics[width=5.75cm]{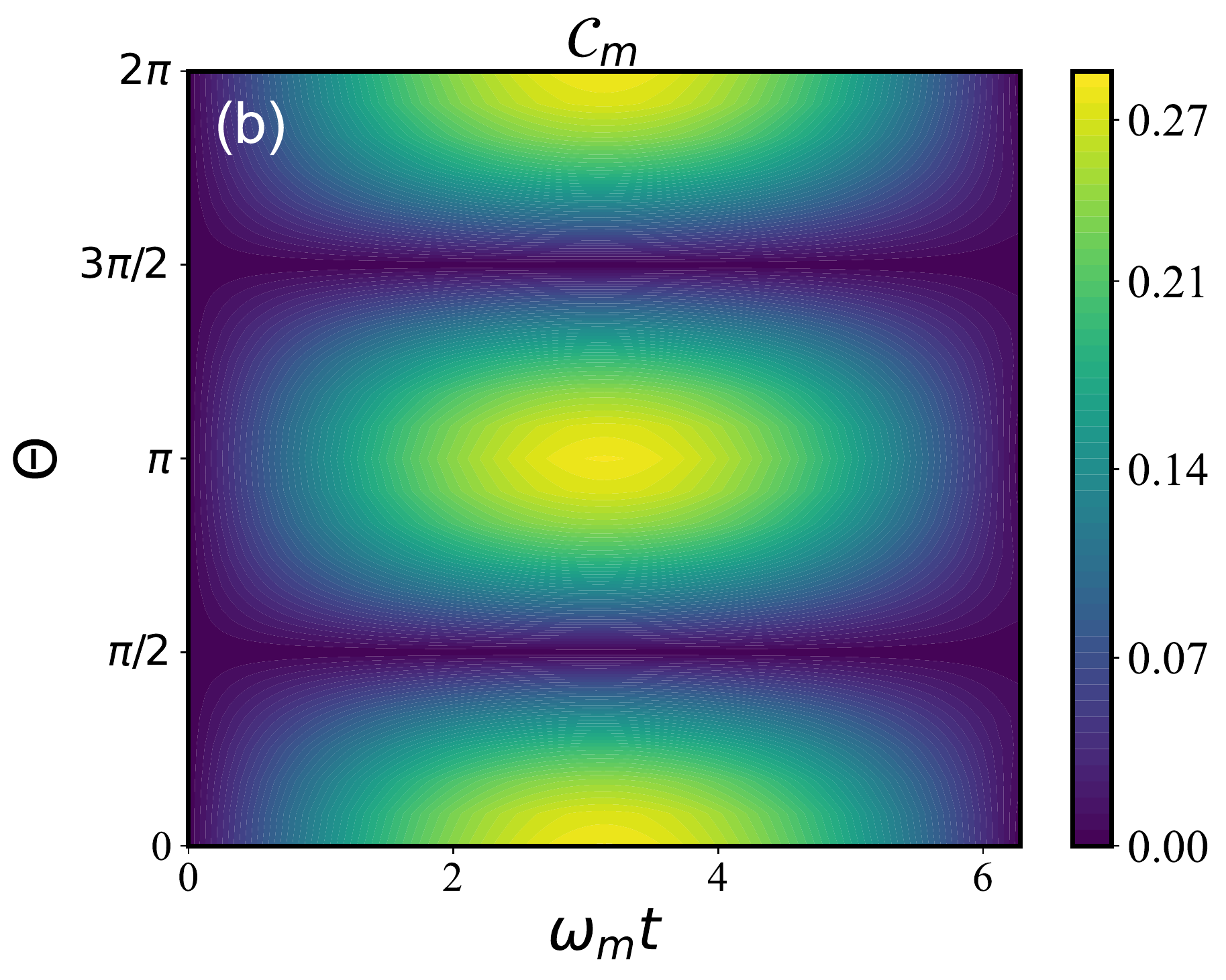}
\hspace*{0.008cm}
\includegraphics[width=5.6cm]{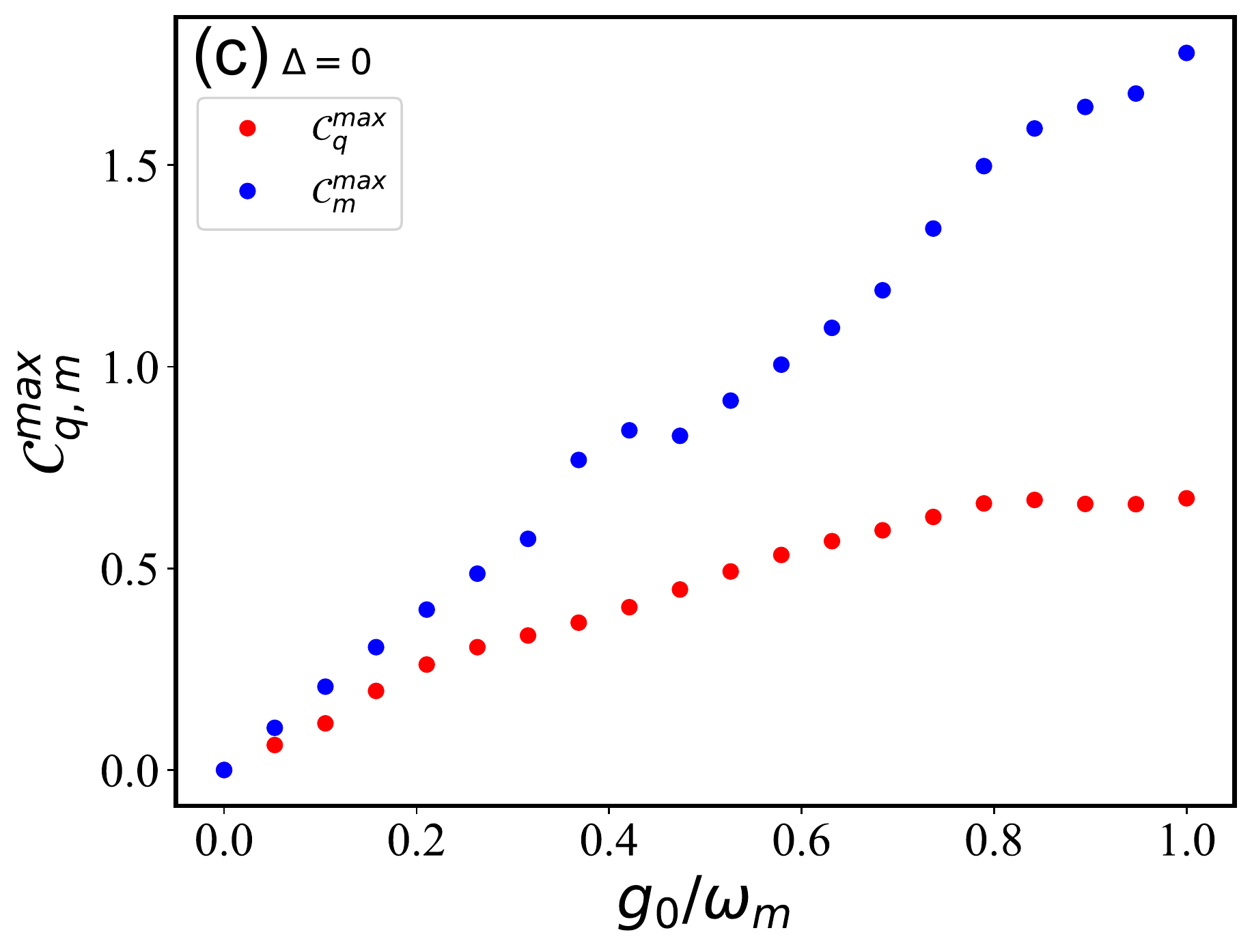}
\\
\includegraphics[width=5.9cm]{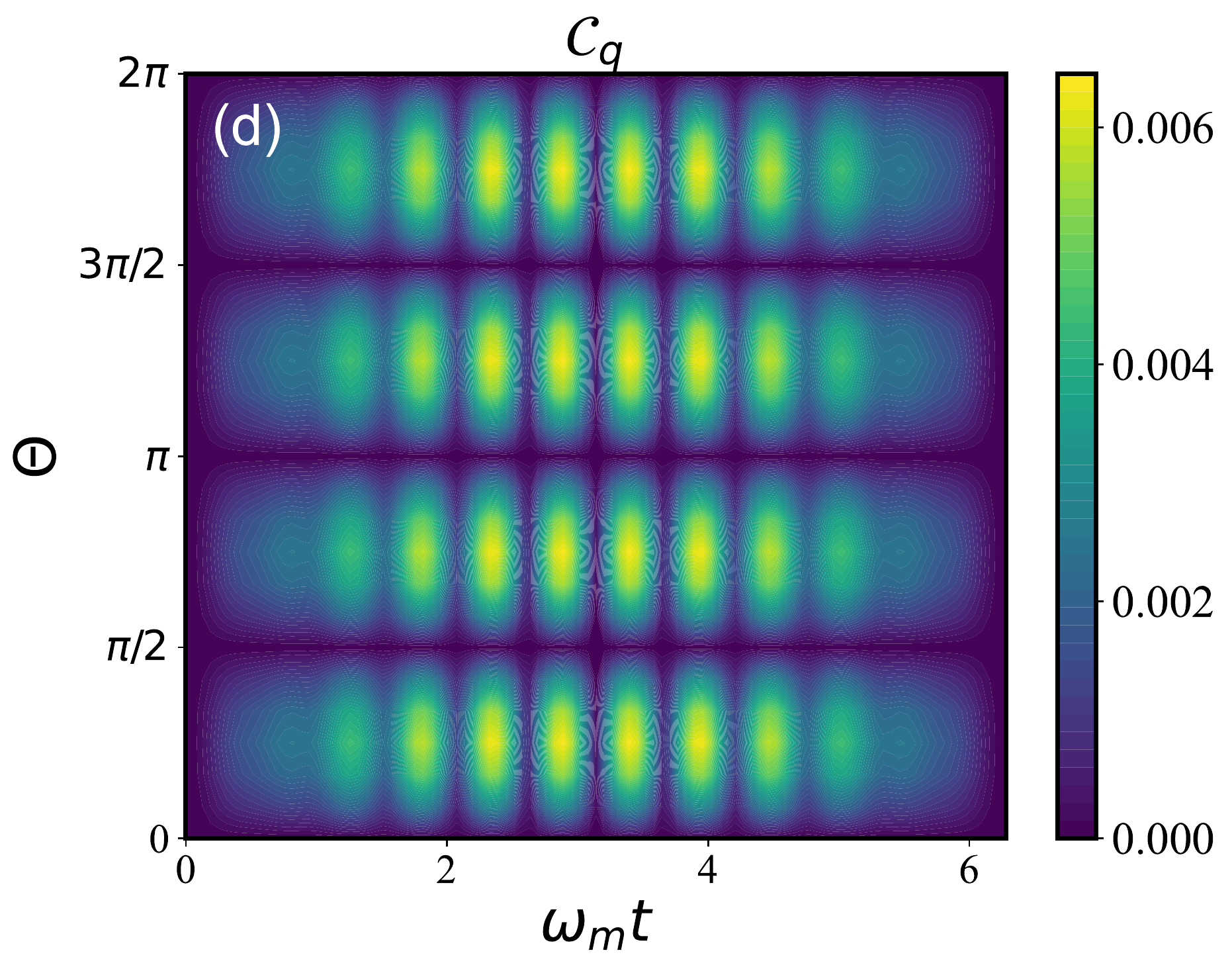}
\hspace*{0.008cm}
\includegraphics[width=5.75cm]{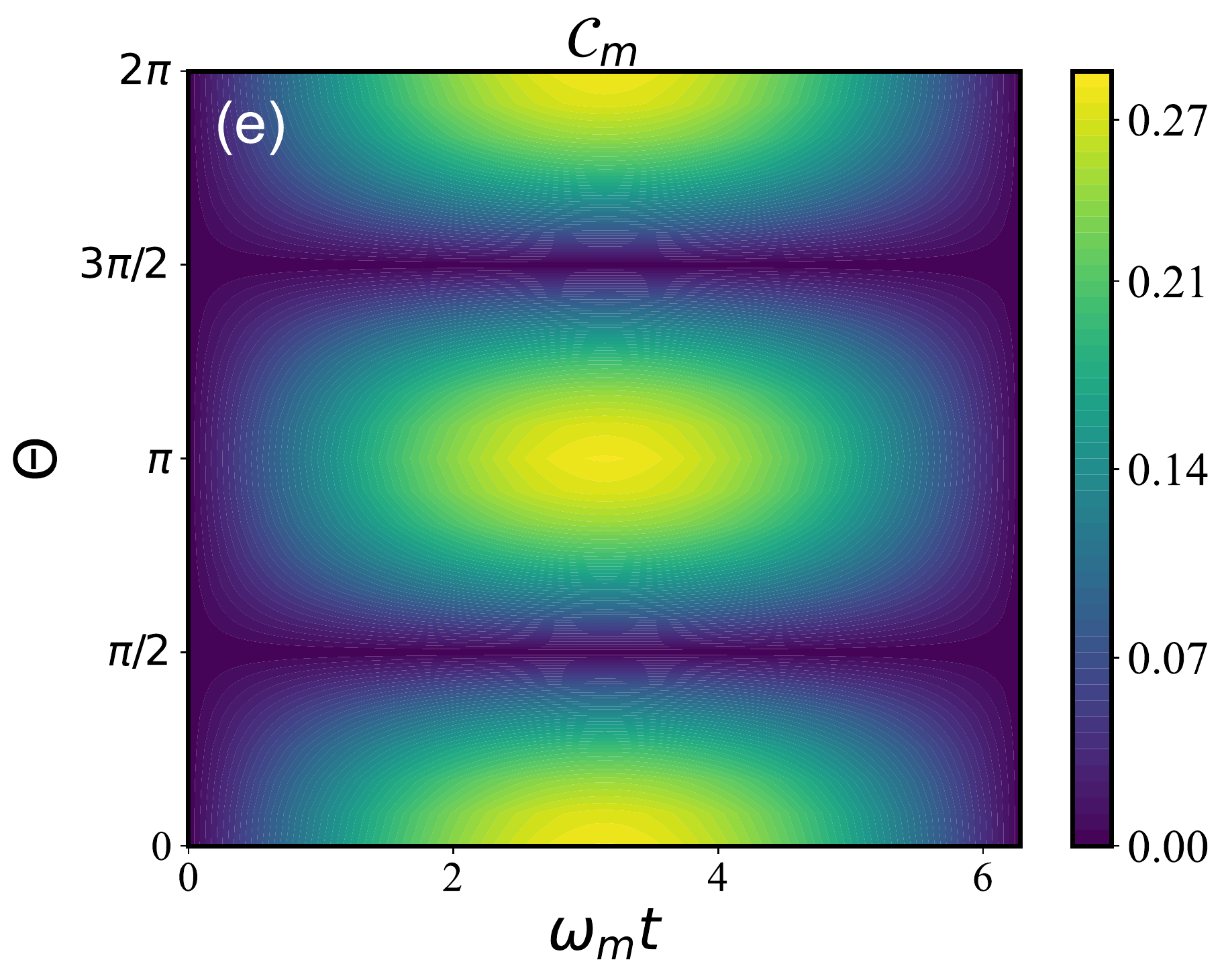}
\hspace*{0.008cm}
\includegraphics[width=5.6cm]{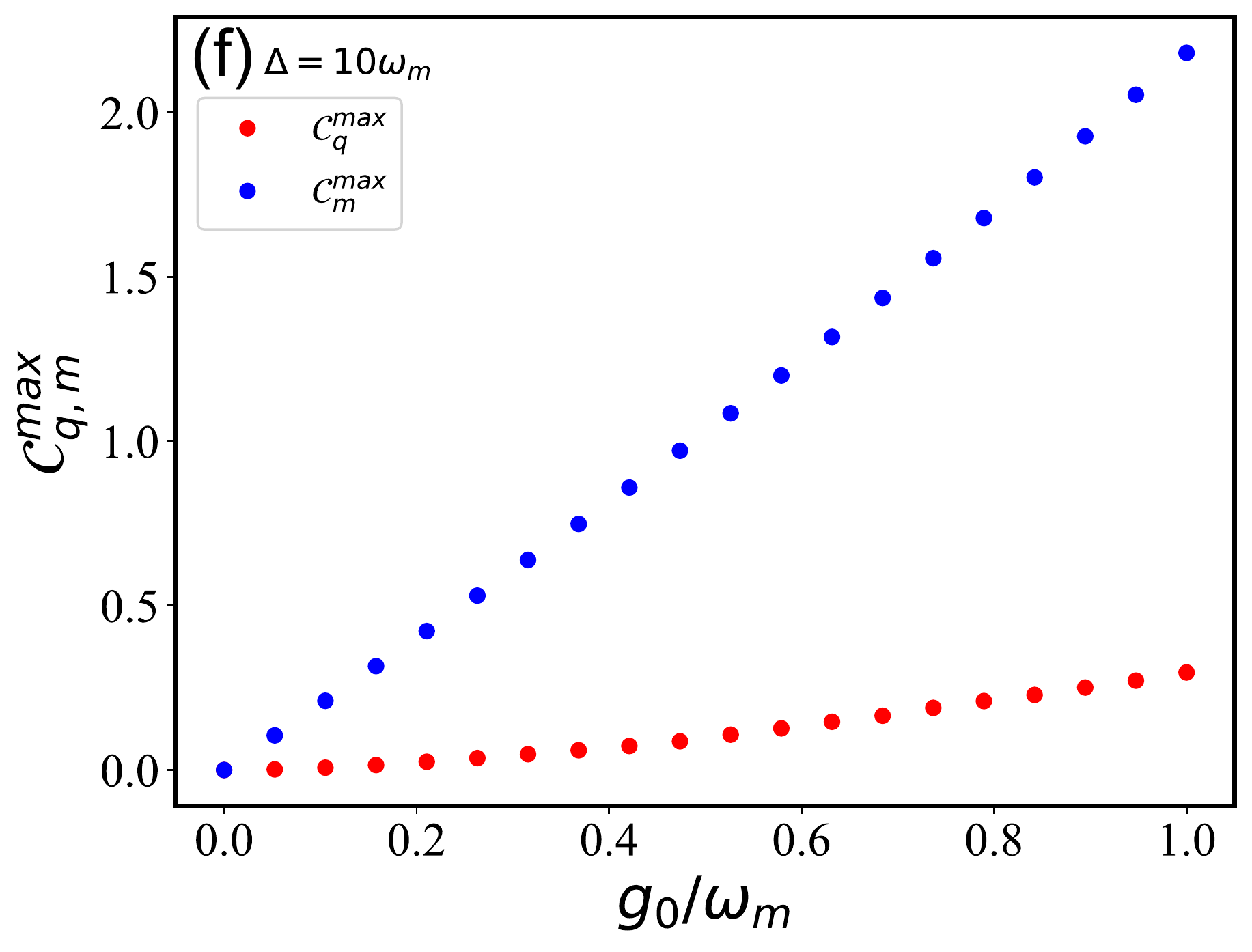}
\caption{Coherence dynamics caused by qubit-oscillator interaction:
  (a,b,d,e) The contour plots of the coherence parameters $\mathcal{C}_q$ and $\mathcal{C}_m$ with respect to the normalized time $\omega_m t$ and $\theta$ when $g_0 = 0.1 \omega_m$. (c,f) The maximal attainable values of the quantum coherence as a function of normalized qubit-mechanical coupling $g_0/\omega_m$ for $\theta = \pi /4$. In (a-c) $\Delta = 0$, in (d-f) $\Delta = 10 \omega_m$. Other parameters are $n_q = 0$ and $n_m = 0.5$: the qubit is initialized in its ground state while the MO is in a thermal state.
    }
	\label{fig3}
\end{figure}
\subsection{Effect of the interaction parameters on the coherence generation}\label{sub3:Eff-coupl-detun} % <<<

To demonstrate the effects of the coupling rates $g_x$ and $g_z$ on the generation of quantum coherence in the system, in first and second columns of Fig.~\ref{fig3}, we showed the evolution of coherence parameters $\mathcal{C}_q$ and $\mathcal{C}_m$ in time and with respect to $\theta$, in weak coupling regime $g_0 = 0.1 \omega_m$, for resonance ($\Delta = 0$) and off-resonance ($\Delta = 10 \omega_m$) conditions, respectively.
For both cases, the maximum oscillator displacement rises at $\omega_m t=\pi$, but the maximum qubit coherence appears delayed at resonance in~\cref{fig3}~(a,b). Out-of-resonance, in~\cref{fig3}~(d,e), both displacement and coherence appear synchronously.

As is seen from Fig.~\ref{fig3}(a,d), the qubit coherence parameter $\mathcal{C}_q(t)$ takes the non-zero value when $\theta \neq n \pi /2 $ ($n = 1,2,\cdots$), i.e., when both $g_x , g_z \neq 0$.  The maximum amount of $\mathcal{C}_q(t)$ can be obtained for $\theta = (2 n +1) \pi /4 $, which shows that $\mathcal{C}_q$ strongly depends on the factor $ |g_x g_z| =  |g_0^2 \sin (2\theta) / 2|$. In addition, increasing the detuning causes a fast reduction in the maximum available amounts of qubit coherence $\mathcal{C}_q$ (compare panels (a) and (d) in Fig.~\ref{fig3}). Moreover, at resonance, the evolution of $\mathcal{C}_q(t)$ becomes maximized around $t \approx 2 m \pi / \omega_m$ ($m \in \mathbb N$), whereas at $\Delta = 10 \omega_m$, the interference pattern shows itself in shorter time interval and the maximum values of $\mathcal{C}_q$ shift to smaller time interval $2 \pi /3 < \omega_m t < 4 \pi /3$.

On the other hand, for the fixed values of $g_0 = 0.1 \omega_m$, $n_m = 0.5$ and $n_q = 0$, the mechanical displacement $\mathcal{C}_m(t)$ is not influenced by changing the detuning (see Fig.~\ref{fig3}(b,e)) and is only affected by the displacement coupling rate $g_z =  g_0 \cos \theta$. Therefore, the maximum amount of $\mathcal{C}_m$ is achieved when $\theta = (2 n + 1) \pi /2$ and $t \approx (2 m-1) \pi / \omega_m$.

In Panels (c) and (f) of Fig.~\ref{fig3}, the maximum values of dynamical coherence parameters are depicted as a function of absolute qubit-mechanical coupling $g_0 / \omega_m$ which shows that the stronger coupling gives rise to higher quantum coherence in the system.

The dependence of the mean values of the mechanical quadratures $\langle {X}_m (t) \rangle$ and $\langle {P}_m (t) \rangle$ on $g_z$ and the Pauli matrices $\langle {\sigma}_x (t) \rangle$ and $\langle{\sigma}_y (t) \rangle$ on both $g_x$ and $g_z$ can also be revealed analytically for a very short time interval in ideal evolution where we can approximate the time evolution operator ${U}(t) = \mathrm{e}^{- i {H} t } \approx \mathbb{I} - i {H} t $. Therefore, the final state of the system up to second order in time  is given by
\begin{equation}\label{rho_f2}
{\rho}_{f}(t) \approx {\rho}(0) - i t \: \Big[{H}, {\rho}(0)\Big] + t^2 {H} \: {\rho}(0) \: {H} + \mathcal O (t^2).
\end{equation}
Under such approximation, the system operators' mean values become
\begin{subequations}\label{xpsxy}
  \begin{align}
    \langle {X}_m (t) \rangle &
    \approx
    \sqrt{ 2 } g_z t^2 \left[ \omega_m n_m (4 n_m +3) (2 P_{ee} - 1) + \frac{\omega_q}{2}(2 n_m +1) \right],
    \\
    \langle {P}_m (t) \rangle &  \approx  - \sqrt{2} g_z t \: (2 P_{ee} - 1), \\
    \langle {\sigma}_x (t) \rangle &  \approx  2 g_x g_z t^2 (2 n_m +1) \left(2 P_{ee} - 1 \right), \label{xpsxy_c}\\
    \langle {\sigma}_y (t) \rangle &  \approx  0. \label{xpsxy_d}
  \end{align}
\end{subequations}
where $P_{n,n} = n_m^n / (1+n_m)^{n+1}$ denotes coefficients of expansion of the initial thermal state of the mechanics in the Fock-state basis, $n_m$ is this state's mean occupation.
From Eq.~\ref{xpsxy}, we see that up to $\mathcal{O}(t^2)$, the mechanical quadratures $\langle {X}_m (t) \rangle$ and $\langle {P}_m (t) \rangle$ are only affected by $g_z$. However, $\langle {\sigma}_x (t) \rangle$ and therefore, $\mathcal{C}_q$ depends on the product $g_x g_z$.
From~\cref{xpsxy_c,xpsxy_d}, we obtain $\Cq \approx  |\langle {\sigma}_x (t) \rangle| = 2 |g_x g_z| t^2 (2n_m+1) / (2 n_q +1)$. This indicates that in short time interval, $\Cq$ changes quadratically with time ($\Cq \propto t^2$).
The qubit coherence $\Cq$ also depends on the mechanical and the qubit occupation number ratio $\Cq \propto (2n_m+1) / (2 n_q + 1)$
which reveals why we could attain better results of $\Cq\up{max}$ when $n_q < n_m$ (see Fig.~\ref{fig2}).
Hence, the best result can be achieved when we fix $n_q = 0$, while increasing the initial occupation $n_m$ (see Fig.~\ref{fig2}).
While the short-time approximation agrees qualitatively with simulations, the quantitative agreement holds only for very short times $\omega_m t \ll 1$.
The maximal values of coherence are reached at considerably longer times which, unfortunately, do not admit the analytical solution.

\begin{figure}[t!]
\centering
\includegraphics[width=7.78cm]{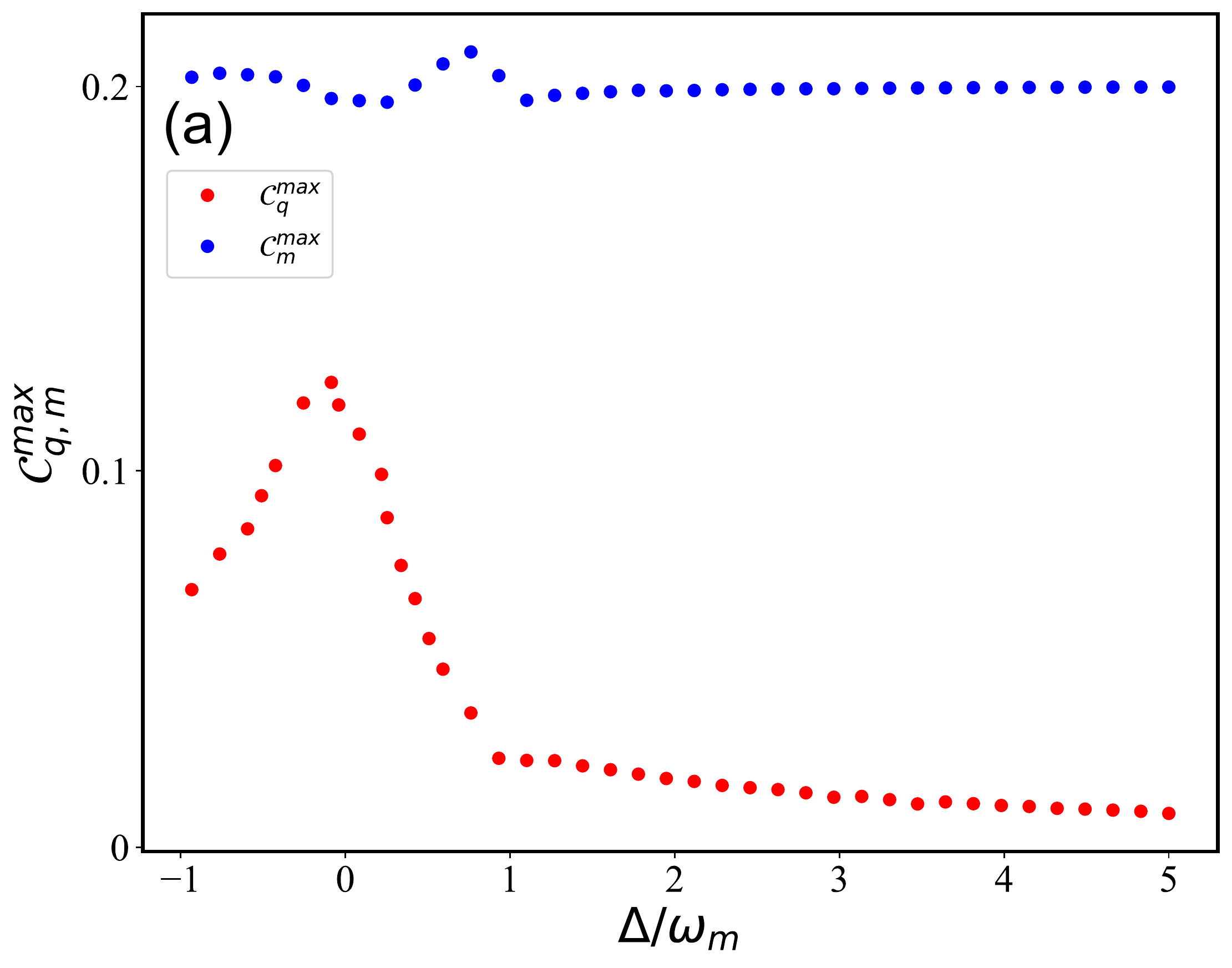}
\hspace*{0.008cm}
\includegraphics[width=8cm]{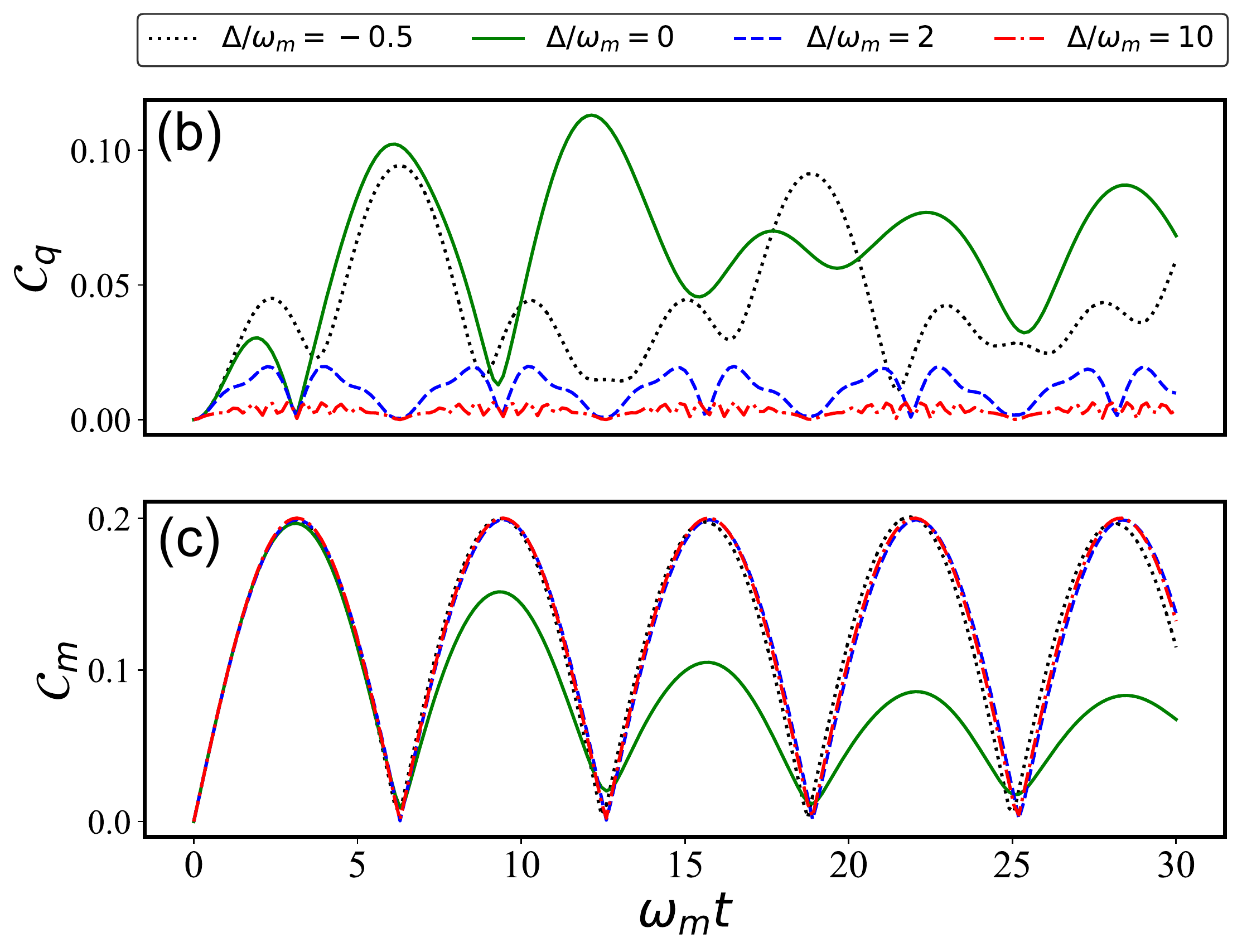}
\caption{Resonant features of emergent quantum coherence: (a) Optimum values of the coherence parameters as a function of the normalized detuning $\Delta / \omega_m$.  The definition of the detuning $\Delta = \omega_q - \omega_m$ does not allow values below $-\omega_m$. The evolution of (b) the qubit coherence $\mathcal{C}_q(t)$ and (c) the mechanical coherent displacement $\mathcal{C}_m(t)$ for different values of detuning. Other numerical parameters are $g_0 = 0.1 \omega_m$, $\theta = \pi/4$, $n_m = 0.5$, and $n_q = 0$.}
	\label{fig4}
\end{figure}

It is also worth looking at the variations of the quantum coherence with respect to the detuning to investigate the resonant nature of this phenomenon. The optimum values of coherence parameters $\mathcal{C}_q\up{max}$ and $\mathcal{C}_m\up{max}$ as a function of normalized detuning $\Delta / \omega_m$ have been demonstrated in Fig.~\ref{fig4}(a), where we can detect a maximum peak for $\mathcal{C}_q\up{max}$ around $\Delta / \omega_m \approx 0 $.
However, the maximum amounts of the mechanical coherent displacement $\mathcal{C}_m\up{max}$ won't alter much as a function of detuning which is consistent with Fig.~\ref{fig3}(b,e) when we fix the values of $g_0 = 0.1 \omega_m$, $\theta = \pi / 4$, $n_m = 0.5$ and $n_q = 0$.

In addition, in panels (b,c) of Fig.~\ref{fig4}, we showed the evolution of $\mathcal{C}_q$  and $\mathcal{C}_m$, respectively, for different values of normalized detuning. As is seen in panel (b), by changing the detuning from $\Delta = -0.5 \omega_m $ to $\Delta = 10 \omega_m $ and moving to the dispersive regime, the amplitude of $\mathcal{C}_q$ diminishes fast.
On the other hand, the oscillation amplitude of $\mathcal{C}_m$ becomes maximized for the initial time interval, and increasing the detuning doesn't change it (see Fig.~\ref{fig4}(c)).

To find out why the coherence parameters respond to the detuning like what is mainly shown in Fig.~\ref{fig4}, it would be better to take a look at the Hamiltonian of the system in the interaction picture, given by
\begin{equation}\label{VI}
{H}^{(I)} = \mathrm{e}^{+i H_0 t} {H} \mathrm{e}^{-i H_0 t} - {H}_0 = g_x ({\sigma}_{-} {a}^{\dagger} \: \mathrm{e}^{-i \Delta t} + {\sigma}_{+} {a} \: \mathrm{e}^{+i \Delta t}) + g_x ({\sigma}_{+} {a}^{\dagger} \: \mathrm{e}^{i \Sigma t} + {\sigma}_{-} {a} \: \mathrm{e}^{-i \Sigma t}) + g_z {\sigma}_z  ({a}^{\dagger} \: \mathrm{e}^{+ i \omega_m t} + {a} \: \mathrm{e}^{-i \omega_m t}),
\end{equation}
where ${a} = ({X}_m + i {P}_m) / \sqrt{2}$ denotes the mechanical annihilation operator, and  $\Sigma = \omega_q + \omega_m$. From Eq.~(\ref{VI}), we can see that for $\Delta \approx 0$, the rotating terms $g_x ({\sigma}_{-} {a}^{\dagger} + {\sigma}_{+} {a} )$, which are responsible for an exchange of excitations between the qubit and the MO, play the dominant role in the dynamics of the system, more specifically in $\mathcal{C}_q$ through the coupling channel $g_x$.
By increasing the absolute value of the detuning, both the rotating and counter-rotating terms in Eq.~(\ref{VI}) start oscillating fast with the frequency of $\Delta$ and $\Sigma$, respectively. In the dispersive regime, where $\Sigma >\Delta \geq 10 \omega_m $, and due to the adiabatic evolution, the energy exchange between the qubit and the MO which mainly happens through the coupling channel $g_x$ diminishes. This affects the qubit coherence which depends on both $g_x$ and $g_z$ factors and leads us to the smaller maximum amounts of $\mathcal{C}_q$. As the mechanical coherent displacement is mainly influenced by the coupling rate $g_z$ and therefore, the displacement term $g_z {\sigma}_z  ({a}^{\dagger} \: \mathrm{e}^{+ i \omega_m t} + {a} \: \mathrm{e}^{-i \omega_m t})$, changing the detuning can not significantly impact $\mathcal{C}_m$ (see Fig.~\ref{fig3}(b,e) and Fig.~\ref{fig4}(a,c)).

\subsection{Quantum coherence in the presence of damping and noise}\label{3B}
\begin{figure}[t!]
\centering
\includegraphics[width=8cm]{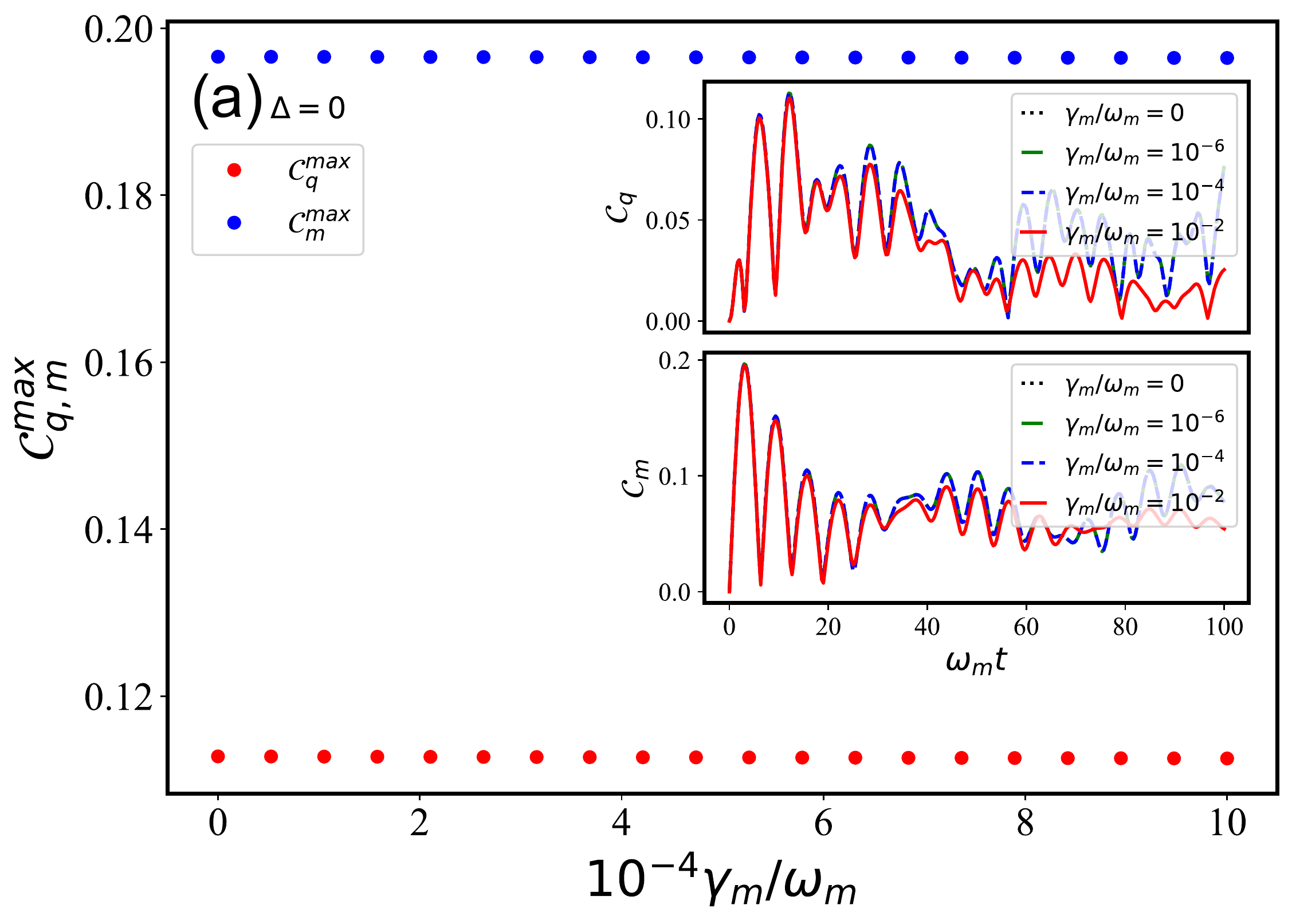}
\hspace*{0.008cm}
\includegraphics[width=8cm]{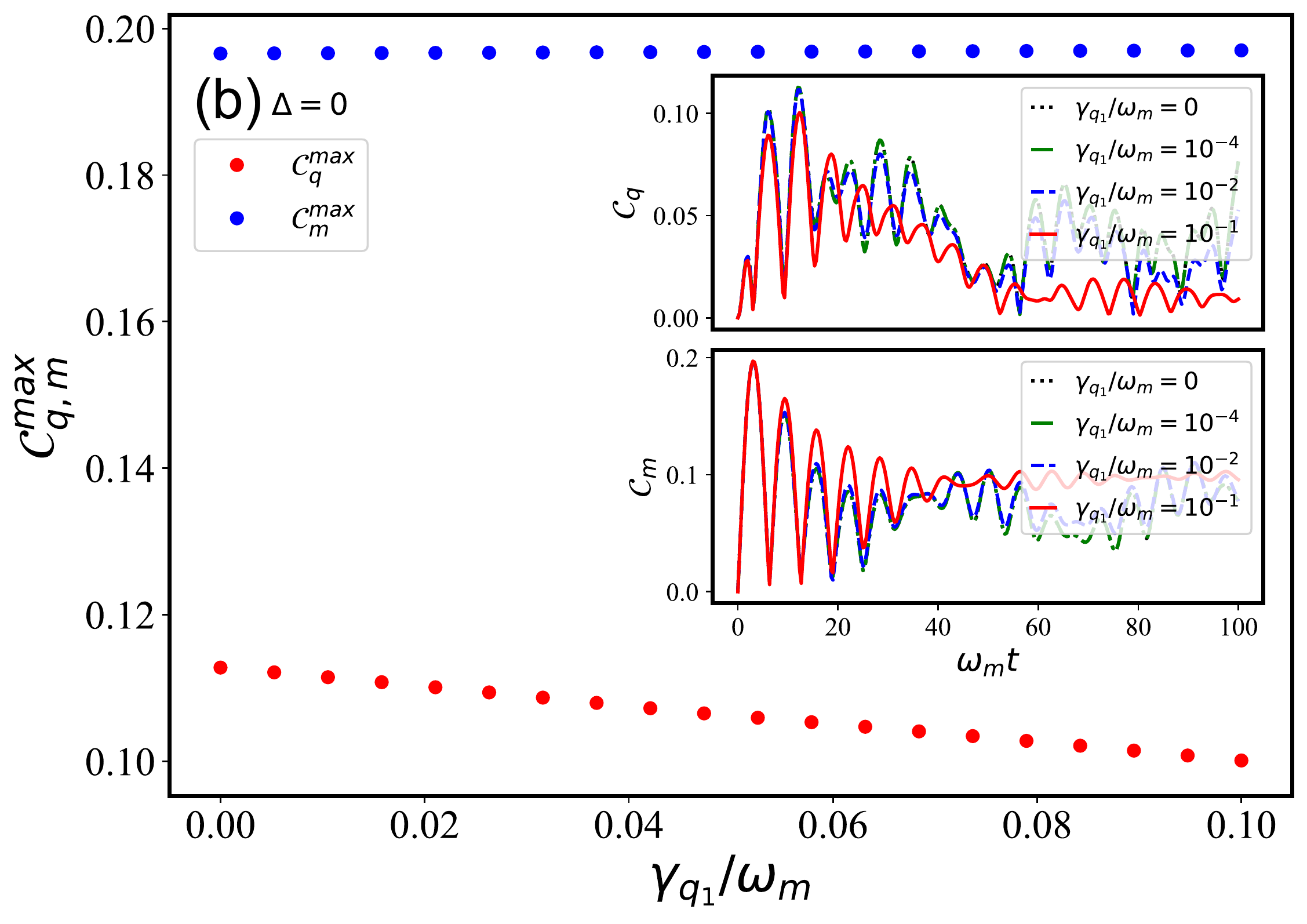}
\caption{Robustness of emerging quantum coherence: optimum values of the coherence parameters as a function of the normalized (a) mechanical damping rate $\gamma_m / \omega_m$ when $\gamma_{q_1} / \omega_m =0$ and (b) qubit damping rate $\gamma_{q_1} / \omega_m$ for $\gamma_m / \omega_m = 10^{-6}$. Inset plots show the evolution of the coherence parameters for different values of (a) $\gamma_m / \omega_m$ and (b) $\gamma_{q_1} / \omega_m$. Other numerical parameters are $\Delta = 0$, $g_0 = 0.1 \omega_m$, $n_m = 0.5$, and $n_q = 0$.
}
	\label{fig5}
\end{figure}
In order to study the dynamics of the system more realistically, we need to take the dissipation and decoherence effects into account.
For an open system interacting with an environment, its density matrix obeys the Lindblad master equation
\begin{equation}\label{LME}
{\dot{\rho}} = - i [{H} , {\rho} ] + \frac{\gamma_m}{2} (n_m + 1) \mathcal{L}({a}) \rho +  \frac{\gamma_m}{2} n_m \mathcal{L}({a}^{\dagger}) \rho + \frac{\gamma_{q_1}}{2} (n_{q} + 1) \mathcal{L}({\sigma}_-) \rho + \frac{\gamma_{q_1}}{2} n_{q} \mathcal{L}({\sigma}_+) \rho
\end{equation}
Here, $\mathcal{L}({O}) = 2 {O} {\rho} {O}^{\dagger} - ({O}^{\dagger} {O} {\rho} +  {\rho} {O}^{\dagger} {O})$ (${O} \equiv {a}, {a}^{\dagger}, {\sigma}_{\pm}$) denotes the Lindblad superoperator.
Further, $\gamma_m$, $\gamma_{q_1} = 1/T_1$ represent the mechanical and qubit relaxation rates, respectively. By solving the master equation (\ref{LME}) numerically, we have investigated the effects of the mechanical and qubit damping on the dynamics of coherence parameters for resonance case $\Delta = 0$ (see Fig.~\ref{fig5}).

In panel (a) of Fig.~\ref{fig5}, we showed the changes of the attainable quantum coherence with respect to the mechanical damping rates $\gamma_m / \omega_m$ in the absence of the qubit dissipation and noise ($\gamma_{q_1} = n_q = 0$) and when the system is operated in resonance condition $\Delta = 0$ and weak coupling regime $g_0 = 0.1 \omega_m$. We also consider the mechanical occupation to be $n_m = 0.5$. As can be seen in Fig.~\ref{fig5}(a), the maximum values of coherence parameters $\mathcal{C}_q\up{max}$ and $\mathcal{C}_m\up{max}$ do not change considerably as $\gamma_m / \omega_m$ increases. In addition, it is evident from the inset plots of Fig.~\ref{fig5}(a) that the dynamical coherence parameters overlap for all $\gamma_m / \omega_m < 10^{-2}$ which means that they are completely robust against mechanical damping as far as $\gamma_m / \omega_m < 10^{-2}$. Moreover, larger values of the mechanical dissipation such as $\gamma_m / \omega_m = 10^{-2}$, do not affect the coherence parameters for the initial time interval (red dotted-dashed lines in inset plots of Fig.~\ref{fig5}(a)). However, we could observe decrease of the coherence parameters for a longer time. By comparing the inset plots in~\ref{fig5}(a), we realize that $\mathcal{C}_q$ decreases with the faster rate than $\mathcal{C}_m$ for $\gamma_m / \omega_m = 10^{-2}$.

The evolution of the coherence parameters in the presence of the normalized qubit damping rate $\gamma_{q_1} / \omega_m$ is plotted in Fig.~\ref{fig5}(b) when we consider $n_m = 0.5$, $n_q = 0$ and $\gamma_m / \omega_m = 10^{-6}$. In this case, we can see that $\mathcal{C}_q\up{max}$ decreases with increasing qubit relaxation rate, while $C_m\up{max}$ does not change much with increasing $\gamma_{q_1} / \omega_m$ which emphasizes the robustness of  mechanical displacement against qubit damping.
The inset plots also confirm these results. In addition, by looking to the inset plots of Fig.~\ref{fig5}(b), it is clear that for $\gamma_{q_1} / \omega_m = 10^{-2}$, coherence parameters would be resistant to the qubit dissipation in shorter time interval $\omega_m t \leq 2 \pi $.
Simulations in the longer time interval show that both $\mathcal{C}_q$ and $\Cm$ decrease and eventually reach small non-zero steady-state values ($\mathcal{C}_q \approx 0.01$, $\Cm \approx 0.1$).

To summarize our study of the influence of the baths, the maximum of attainable coherence seems to be reached at rather early times that amount to the interaction running for only a few periods of mechanical oscillations.
For the state-of-the-art electromechanical systems, due to their exceptional Q-factors, the interaction at these timescales is very close to unitary.
Therefore, we can state that the interaction with thermal reservoirs during the coherent interaction between the mechanics and the qubit, has very limited effect on the maximal coherence attainable from fast pulsed interaction studied here.

\section{Discussion}

In this article, in contrast to the previous steady-state studies~\cite{guarnieri_steadystate_2018,roman-ancheyta_enhanced_2021} we theoretically investigated the possibility of generating transient quantum coherence in a qubit-mechanical system from incoherent thermal states.
We studied the transient interaction between a charge qubit and a mechanical oscillator, similar to what is found in electromechanical setups~\cite{ma_nonclassical_2021,lahaye_nanomechanical_2009,oconnell_quantum_2010,rouxinol_measurements_2016,chu_quantum_2017,sletten_resolving_2019,wollack_quantum_2022}.
We showed how the sensitivity of the qubit to the offset charge enables us to couple the qubit to the mechanical motion in both vertical and parallel ways with respect to the eigenstates of the free Hamiltonian of the qubit as far as the system is operated near the degeneracy point.
The simultaneous presence of these two different coupling rates allows the observation of the qubit coherence in the system with the initial incoherent thermal state.
This is so in both the ideal case of unitary interaction and in the dissipative situation.
It should be noted that in this model, dynamical coherence emerges without the use of conventional methods such as coherent driving~\cite{bloch_nuclear_1946} or coherence measurement~\cite{cable_measurementinduced_2005,filip_thermally_2014}.

Differently to the steady-state coherence, the thermal occupation number of the mechanical mode has a positive effect on generating larger coherence of the qubit.
We observed that increasing the net values of the coupling rate $g_0$ causes an improvement in the maximum accessible amounts of the qubit coherence and mechanical coherent displacement.
In addition, we demonstrated how the parallel and perpendicular components $g_z$ and $g_x$ of the coupling rates affect the quantum coherence. In the case of the qubit coherence $ \mathcal{C}_q$, the product $g_x g_z$ plays the main role while for mechanical displacement $\mathcal{C}_m$, the parallel coupling $g_z$ becomes important. The maximum coherence values for the qubit and the MO could be obtained for $|g_x| = |g_z|$, i.e., when we set the optimum value $\theta = \pi /4$ for the coupling phase.
Moreover, we showed that the qubit coherence parameter is strongly dependent on the detuning $ \Delta $ through the coupling channel $g_x$ such that by adjusting the detuning and setting it close to resonance $ \Delta \approx 0$, where the role of the rotating term associated with the coupling rate $g_x$ gets dominant, we reach the maximum values for the qubit coherence parameter. However, changing the detuning can not significantly alter the maximum values of mechanical displacement.
Finally, we obtained that the mechanical coherence generated in our model is almost robust against both the mechanical and the qubit damping processes, while the larger values of qubit damping rate ($ \gamma_{q_1} > 10^{-2} \omega_m$) give rise to the decaying of the qubit coherence parameter.

The experimental realization of such a model has been already demonstrated in Ref.~\cite{ma_nonclassical_2021}. Aside from the electromechanical setups, there are other experimental platforms for the realization of our model such as trapped ions~\cite{lo_spinmotion_2015,kienzler_observation_2016} and NV-centers coupled magnetically to the mechanical motion~\cite{rabl_strong_2009,kolkowitz_coherent_2012,li_coupling_2022}. Hybrid atom-optomechanical and electro-optomechanical systems also provide a great potential for this purpose~\cite{dantan_hybrid_2014,rogers_hybrid_2014,abari_generation_2020,dong_unconventional_2021,nongthombam_groundstate_2021,arguello-luengo_optomechanical_2021}.

Quantum coherence counts among fundamental resources in quantum information processing and quantum computation~\cite{streltsov_colloquium_2017,winter_operational_2016,yogesh_quantum_2022}.
It also provides great applications in the context of quantum sensing~\cite{malinovskaya_atomic_2015}, quantum thermodynamics~\cite{lostaglio_description_2015,horodecki_fundamental_2013,korzekwa_extraction_2016}, quantum biology~\cite{lloyd_quantum_2011} and non-equilibrium models~\cite{francica_role_2019,santos_role_2019,vanvu_finitetime_2022}.
In each of these fields, autonomous emergence of quantum coherence can be beneficial.
Such proof-of-principle experimental tests will further investigate the emergence of quantum coherence and extensions of the mechanisms we addressed here.

\section{Methods}\label{3C}\label{sec:methods}

\subsection{Tools for numerical calculation}\label{3C1}
In this manuscript, we use the QuTiP package~\cite{johansson_qutip_2013,johansson_qutip_2012} to numerically investigate the evolved density matrix as well as coherent properties of the system in both ideal and dissipative situations.
For the ideal case, we solve the von Neumann equation $\dot{{\rho}}(t) = - i [{H} , {\rho}]$ with the initial condition (\ref{rho_i}). However, the total density matrix in an open system is obtained by solving the master equation
\begin{equation}\label{rhofM}
{\dot{\rho}} = - i [{H} , {\rho} ] + \sum_n \frac{1}{2} \left( 2 {\mathcal{A}}_n {\rho}(t) {\mathcal{A}}_n^{\dagger} -  {\rho}(t) {\mathcal{A}}_n^{\dagger} {\mathcal{A}}_n - {\mathcal{A}}_n^{\dagger} {\mathcal{A}}_n  {\rho}(t) \right),
\end{equation}
numerically, where in our system ${\mathcal{A}}_1 = \sqrt{\gamma_m (n_m + 1)} \: {a}$, ${\mathcal{A}}_2 = \sqrt{\gamma_m n_m} \: {a}^{\dagger}$, ${\mathcal{A}}_3 = \sqrt{\gamma_{q_1} (n_{q} + 1)} \: {\sigma}_-$ and ${\mathcal{A}}_4 = \sqrt{\gamma_{q_1} n_{q}} \: {\sigma}_+$.
As mentioned before, the Hamiltonian ${H}$ appearing in von Neumann and master equations is given by
\begin{equation}\label{H00}
{H} = {H}_0  + {H}_{int},
\end{equation}
where ${H}_0 = {H}_q + {H}_m$ characterizes the free dynamics of the qubit and the MO with ${H}_q =  \omega_q {\sigma}_z /2$ and $ {H}_m = \omega_m({X}_m^2 + {P}_m^2) /2 $.
The general form of the interaction term between a qubit and the MO can be modeled as
\begin{equation}\label{Hint1}
{H}_{int} = g_0 \, (\mathbf{n} \cdot \vec{{\sigma}}) {X}_m,
\end{equation}
where $\mathbf{n}$ is a normal vector in Bloch space such that
\begin{equation}\label{nsigma}
\mathbf{n} \cdot \vec{{\sigma}} = {\sigma}_x \cos \phi \sin \theta + {\sigma}_y \sin \phi \sin \theta + {\sigma}_z \cos \theta.
\end{equation}
In most experimental works~\cite{oconnell_quantum_2010,wollack_quantum_2022}, the mechanical mode only couples to the one component of the Pauli matrix, i.e.,  $ {\sigma}_x {X}_m$ ($\phi = 0$, $\theta = \pi /2$). However, it is also possible to couple the mechanical motion to more than one component of the Pauli matrix due to the imperfection of the quantum circuit.
An experimental realization of such model can be achieved by an electromechanical system, where a nanomechanical oscillator coupled capacitively to a Cooper-pair box (CPB) as a charge qubit operating near the so-called degeneracy point (see Fig.~\ref{fig1}(b))~\cite{ma_nonclassical_2021}. In this setup, the tiny vibration of the mechanical oscillator can modify the gate-voltage $V_g ({x})$ as well as the gate-capacitor $C_g ({x})$ such that the gate-charge $n_g({x}) = C_g({x}) V_g({x}) / 2e$ becomes mechanically position-dependent (see Appendix~\ref{A1}).
By controlling the sensitivity of the charge qubit with respect to the gate charge $n_g({x})$, the direct coupling between the qubit and the MO becomes possible.

For the charge qubit, the dynamics and the transition frequency $\omega_q$ are strongly dependent on gate-charge $n_g({x})$ and therefore on mechanical displacement operator ${x}$. Such dependence on the one hand could be destructive as the offset-charge can induce noise to the qubit and increases its decoherence rate. On the other hand, it induces a desirable coupling between the qubit and the mechanical modes in our model.
In this case, ${H}_{int} = \left( g_x \, {\sigma}_x + g_z \, {\sigma}_z \right) {X}_m$ describes the interaction Hamiltonian
where $g_y = 0 $ (for $\phi=0$) and $g_x = g_0 \sin \theta $, while $g_z = g_0 \cos \theta $ characterizes the residual coupling rate (see Appendix~\ref{A1}).

The presence of the coupling term $ g_x \, {\sigma}_x {X}_m$ and the additional coupling $ g_z \, {\sigma}_z {X}_m$ at the same time, which contain the perpendicular and parallel components ${\sigma}_x$ and ${\sigma}_z$, with respect to the free Hamiltonian of the qubit ${H}_q = \omega_q {\sigma}_z /2$, make it possible to produce a coherent state for a qubit from the completely incoherent initial state (\ref{rho_i}).
In addition, the presence of an additional term $ g_z \, {\sigma}_z {X}_m$ in this case, which also contains mechanical displacement, applies the net average force on the MO. This allows the observation of the mechanical coherence in the system as well.

To quantify the quantum coherence of the qubit and the MO, we employ the measure of the $l_1$-norm of coherence and define the qubit coherence as $\mathcal{C}_q(t)   =  \sqrt{\langle {\sigma}_x (t) \rangle^2 +  \langle {\sigma}_y (t) \rangle^2}$ and use $\mathcal{C}_m(t)  =  \sqrt{\langle {X}_m (t) \rangle^2 +  \langle {P}_m (t) \rangle^2}$ for the mechanical coherent displacement, respectively.
The expectation values of time-dependent operators $\langle {\sigma}_x (t) \rangle $, $\langle {\sigma}_y (t) \rangle$, $\langle {X}_m (t) \rangle$ and $\langle {P}_m (t) \rangle $ are determined through the following relations
\begin{subequations}\label{mean1}
\begin{align}
\langle {\sigma}_{x(y)}(t) \rangle  & =  \mathrm{Tr} \left[{\rho}(t) \, ({\sigma}_{x(y)} \otimes \mathbb{I}_n )\right] = \mathrm{Tr} \left[{\rho}_q(t) \, {\sigma}_{x(y)} \right], \label{meanSxy} \\
\langle {X}_m(t) \rangle  & =  \mathrm{Tr} \left[{\rho}(t) \, ( \mathbb{I}_q \otimes {X}_m) \right] = \mathrm{Tr} \left[{\rho}_{m}(t) \, {X}_m \right], \label{meanXm} \\
\langle {P}_m(t) \rangle  & =  \mathrm{Tr} \left[{\rho}(t) \, (\mathbb{I}_q \otimes {P}_m ) \right] = \mathrm{Tr} \left[{\rho}_{m}(t) \, {P}_m \right], \label{meanPm}
\end{align}
\end{subequations}
where $\mathbb{I}_n$, $\mathbb{I}_q$ are the identity operators for the qubit and the MO, ${\rho}(t)$ represents the evolved density matrix of the system, while ${\rho}_{q}(t)$ and ${\rho}_{m}(t)$ denote the reduced density matrices of the qubit and the MO, respectively.
Once we compute the evolved density matrix of the system in both ideal and non-ideal situations, we can easily calculate the coherence parameters.

\begin{acknowledgments}
N.E.A. acknowledges the project CZ.02.1.01/0.0/0.0/16\_026/0008460 of MEYS CR.
A.A.R. and R.F. acknowledge the support of the project 20-16577S of the Czech Science Foundation.
R.F. also acknowledges the grant  LTAUSA19099 of MEYS CR.
\end{acknowledgments}

\appendix
\section{Extracting the interaction Hamiltonian of the qubit-mechanical system}\label{A1}

\begin{figure}[t!]
\centering
\includegraphics[width=10cm]{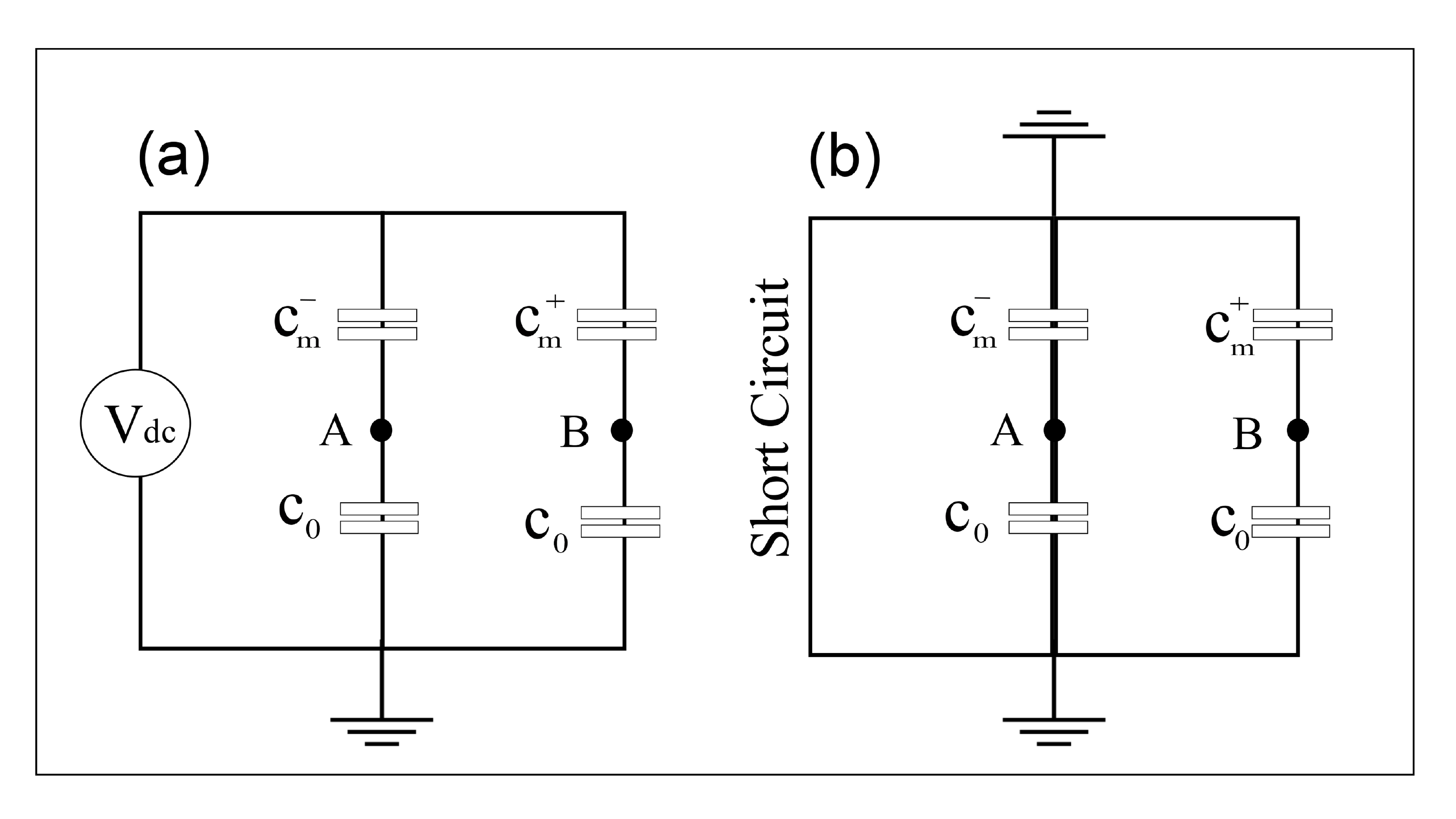}
	\caption{ (a) The Thevenin equivalent representation of the circuit in Fig.~\ref{fig1}(c,d) for calculating $V_g({x})$, and (b) the equivalent short circuit used for calculating $C_g ({x})$.
          }
	\label{figA}
\end{figure}

To extract the interaction term, we start with the equivalent circuit of Fig.~\ref{fig1}(d), such that the equivalent voltage $V_g({x})$, which is the voltage difference across open terminals A and B (the equivalent voltage applied across the Josephson junction), is given by (see Fig~\ref{figA}(a))
\begin{equation}\label{Vgx}
V_g({x}) = V_A ({x}) - V_B ({x}) = V_{\mathrm{dc}} \left( \frac{C_m^{-}({x})}{C_m^{-}({x})+C_0} -\frac{C_m^{+}({x})}{C_m^{+}({x}) + C_0} \right),
\end{equation}
with
\begin{equation}\label{Cmpm}
C_m^{\pm} ({x}) = \frac{\epsilon_0 A}{x_0 \pm {x}} = \frac{C_m^0}{(1 \pm {x} / x_0)},
\end{equation}
where $x_0$ indicates the static separation between the parallel plate capacitors $C_m^{\pm} ({x})$, while $\epsilon_0$ and $A$ represent the permittivity and area of the plate capacitors, respectively. By expanding $V_g({x})$ around small motion at $x=0$, we have
\begin{equation}\label{Cmpm2}
C_m^{\pm} ({x}) \approx  C_m^0 (1 \mp \frac{{x}}{x_0}),
\end{equation}
\begin{equation}\label{Vg2}
V_g({x}) \approx  2 V_{\mathrm{dc}} \frac{C_m^0 C_0}{(C_m^0 + C_0)^2} \cdot \frac{{x}}{x_0} + \mathcal{O} \left( \frac{{x}}{x_0} \right)^2.
\end{equation}
Similarly, the equivalent capacitance $C_g({x})$ is found by replacing the DC-voltage source with a short circuit (Fig.~\ref{figA} (b)),
\begin{equation}\label{Cg1}
\frac{1}{C_g ({x})} = \frac{1}{C_{eq} ({x})} = \frac{1}{C_0 + C_m^{-}({x})} + \frac{1}{C_0 + C_m^{+}({x})},
\end{equation}
\begin{equation}\label{Cg2}
C_g ({x}) \approx  \frac{1}{2} (C_0 + C_m^0) +\mathcal{O}\left( \frac{{x}}{x_0} \right)^2.
\end{equation}
Up to the first order in ${x}$, only the gate-voltage is linearly controlled by the mechanical displacement. Therefore, the off-set charge $n_g (x) = C_g(x) V_g(x) /2$ becomes
\begin{equation}\label{ng2}
n_g ({x}) \approx  \frac{V_{\mathrm{dc}}}{2 e x_0} \cdot \frac{C_m^0 C_0}{(C_0 + C_m^0)} {x}.
\end{equation}
The general Hamiltonian of the qubit in the presence of the mechanical motion is given by
\begin{equation}\label{Hq}
{H}_q ({x}) = 4 E_c \left({n}-n_g({x}) \right)^2 - E_J \cos{{\varphi}},
\end{equation}
where $E_c$ and $E_J$ are the charging and Josephson energy, respectively,
${n}$ is the Cooper-pair number operator and ${\varphi}$ is the superconducting phase operator which can be related to the flux operator through ${\varphi} = 2 \pi {\Phi} / \Phi_0 $,
where $\Phi_0 = h/(2e)$ is the flux quantum. In the number-operator basis the second term of Eq.~(\ref{Hq}) can be written as
\begin{equation}\label{2nd}
 - E_J \cos{{\varphi}} = - \frac{E_J}{2} \sum_n \Big( \vert n \rangle \langle n+1 \vert + \vert n+1 \rangle \langle n \vert \Big).
\end{equation}
The eigenenergies of the Hamiltonian~(\ref{Hq}) for each n-subspace is given by
\begin{equation}\label{En}
\lambda_{\pm}^{(n)}({x}) = 4 E_c \left(n-n_g({x}) \right)^2 + 4 E_c \left(n-n_g({x})\right)+ 2 E_c \pm \frac{1}{2} \sqrt{E_J^2 + (4 E_c)^2 \left(1+2 n - 2 n_g({x})\right)^2}.
\end{equation}
Taking the lowest two energy-levels $\vert n=0 \rangle$ and $\vert n=1\rangle$ as a ground and excited states of a qubit, respectively, into account, the qubit frequency becomes
\begin{equation}\label{flux}
\omega_q ({x}) = \omega_q^{(0)} ({x}) = \sqrt{E_J^2+(4 E_c)^2 \left(1-2 n_g({x})\right)^2},
\end{equation}
and the Hamiltonian~(\ref{Hq}) takes the following form
\begin{equation}\label{Hq2}
{H}_q({x}) \approx  4 E_c \left(1 - 2 n_g({x}) \right) \vert 1 \rangle \langle 1 \vert + 4 E_c n_g^2({x}) \: \mathbb{I} -\frac{E_J}{2} \left( \vert 0 \rangle \langle 1 \vert + \vert 1 \rangle \langle 0 \vert \right).
\end{equation}
Now, the interaction Hamiltonian near the charge degeneracy point $n_g \approx 1/2$ is given by,
\begin{equation}\label{HI}
{H}_{int} = \frac{\partial {H}_q}{\partial x} {x} \Big{\vert}_{n_g \rightarrow \frac{1}{2}} = 8 E_c ({n} - n_g) \frac{\partial n_g ({x})}{\partial x} {x} \Big{\vert}_{n_g \rightarrow \frac{1}{2}} \approx 8 E_c \Big[ \vert 1 \rangle \langle 1 \vert - n_g \Big] \frac{\partial n_g ({x})}{\partial x} {x} \Big{\vert}_{n_g \rightarrow \frac{1}{2}}
\end{equation}
where
\begin{equation}\label{ng3}
\frac{\partial n_g ({x})}{\partial x} = \frac{V_{\mathrm{dc}}}{2 e x_0} \cdot \frac{C_m^0 C_0}{(C_0 + C_m^0)}.
\end{equation}
Using the diagonal bases
\begin{align}\label{ds}
    \vert + \rangle  & =  \cos \vartheta \vert 0 \rangle + \sin \vartheta \vert 1 \rangle, \\
    \vert - \rangle  & =  - \sin \vartheta \vert 0 \rangle + \cos \vartheta \vert 1 \rangle, \nonumber
\end{align}
where $2 \vartheta = \pi / 2 - \theta_0 $ and $\theta_0 = \arctan [4E_c (1-2n_g) /  E_J ] $~\cite{ma_nonclassical_2021}, the Eq.~(\ref{HI}) can be written as
\begin{equation}\label{HI2}
{H}_{int}  =   g_0 {X}_m \left[ \cos \theta_0 {\sigma}_x - \sin \theta_0 {\sigma}_z  +  (1 - 2 n_g) \right] \Big{\vert}_{n_g \rightarrow \frac{1}{2}}.
\end{equation}
Here, we define ${x} = \sqrt{2} x_{\mathrm{zpf}} \: {X}_m$ with $x_{\mathrm{zpf}} = \sqrt{ \hbar / (2 m \omega_m)}$ being the zero-point fluctuation,
 ${\sigma}_z \equiv \vert + \rangle \langle + \vert - \vert - \rangle \langle - \vert$ and ${\sigma}_x \equiv \vert + \rangle \langle - \vert + \vert - \rangle \langle + \vert$ denote the $z$ and $x$ components of Pauli matrix, aligned with the energy quantization axis and perpendicular to it, respectively, and $g_0$ is the single phonon qubit-mechanics coupling, defined as
\begin{equation}\label{gm}
g_0 = \frac{4 E_c}{2 e }\cdot \frac{C_m^0 C_0}{(C_0 + C_m^0)}\cdot \frac{x_{\mathrm{zpf}}}{ x_0} V_{\mathrm{dc}}.
\end{equation}
By using the spherical coordinates in the Bloch space, where $\phi = 0$ and $\theta \equiv \theta_0 + \pi/2 $, the general form of interaction Hamiltonian (\ref{eq:hamiltonian_first}) is derived and the qubit-mechanical coupling rates $g_x$ and $g_z$ can be extracted as
\begin{align}\label{gg}
    g_x  & =  g_0 \cos \theta_0 = g_0 \sin \theta ,  \\
    g_z  & =  - g_0 \sin \theta_0  = g_0 \cos \theta, \nonumber
\end{align}
In addition, standing close to the degeneracy point induces a small qubit-independent shift (QID) with the coupling rate $g_m = g_0 (1- 2 n_g)$ to the MO which is negligible for $n_g \rightarrow 1/2$ such that we have ignored it in Eq.~(\ref{eq:hamiltonian_first}).
The complete dynamical behavior of the coherence parameters in the presence of this shift has been discussed in Appendix~\ref{A2}.

\begin{figure}[t!]
\centering
\includegraphics[width=6.05cm]{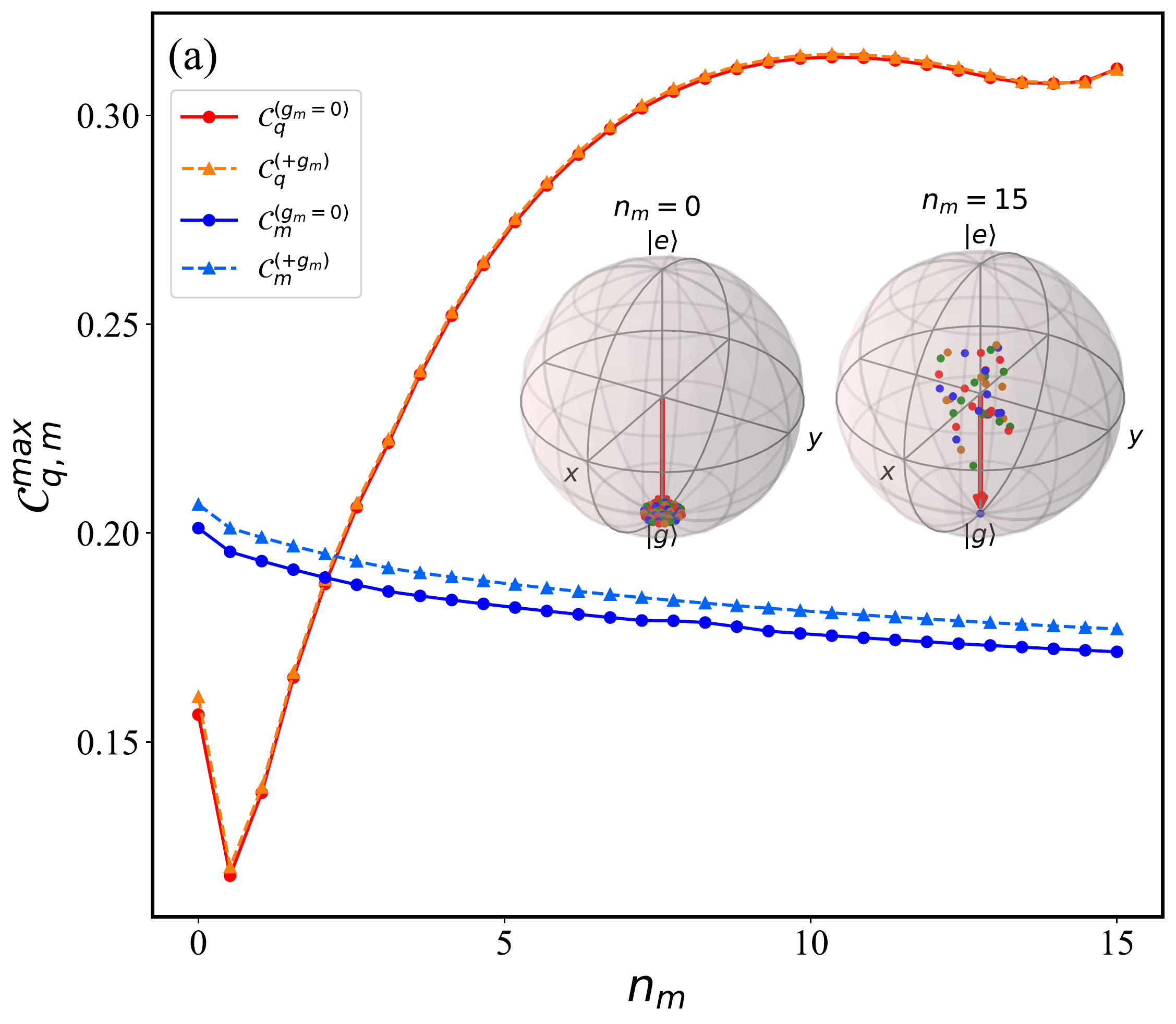}
\hspace*{0.0008cm}
\includegraphics[width=6.05cm]{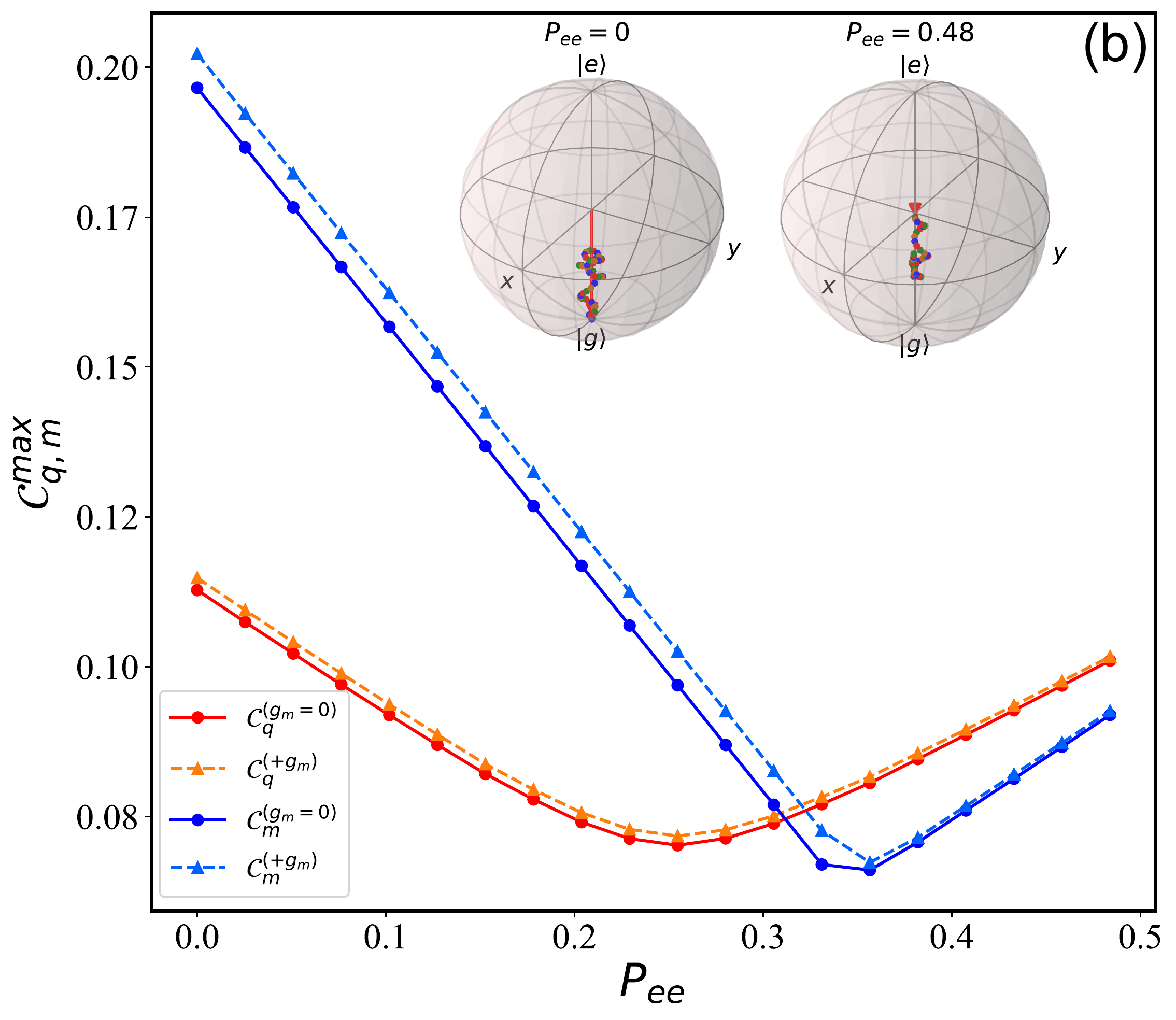}
\hspace*{0.0008cm}
\includegraphics[width=5.4cm]{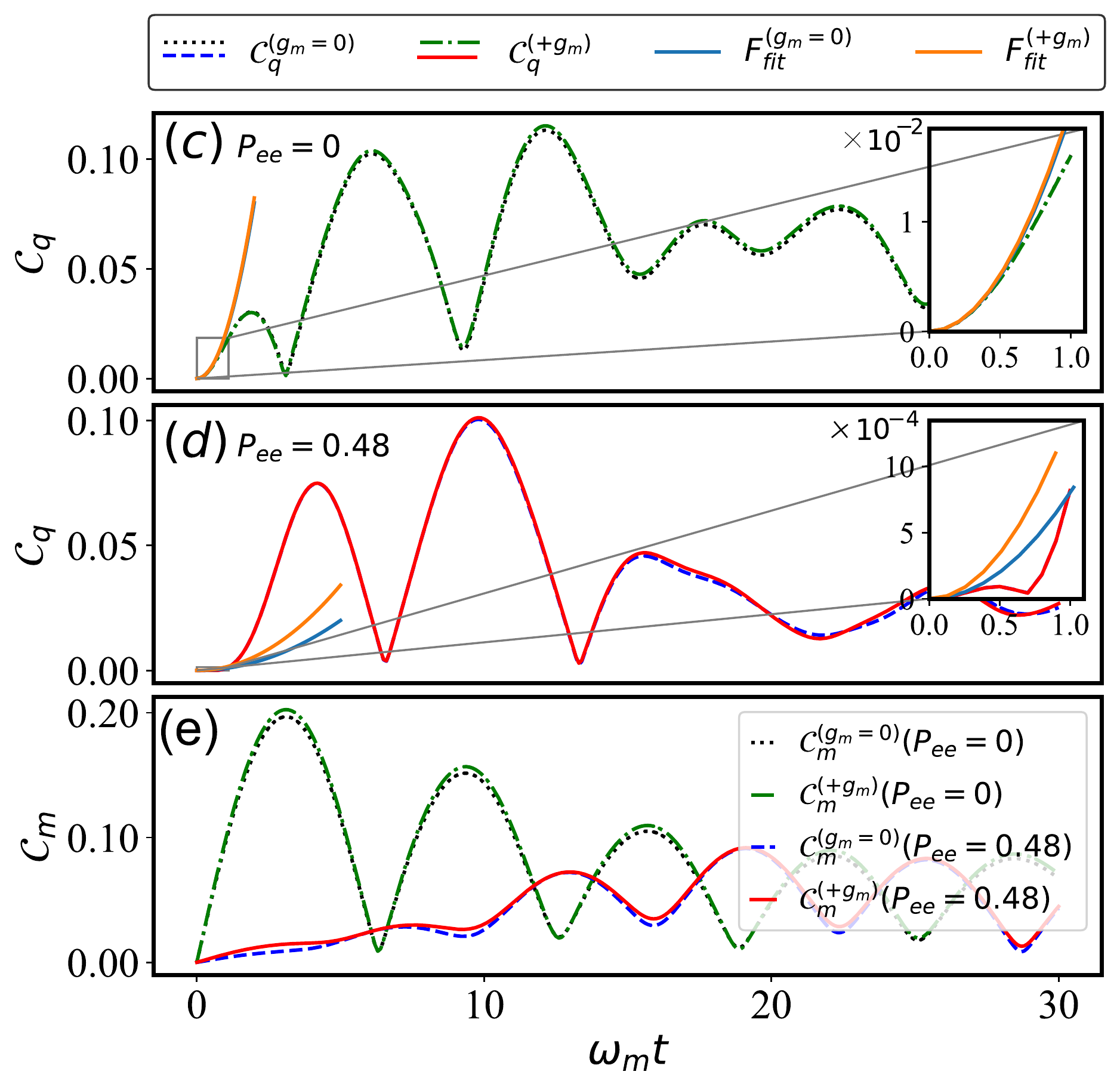}
	\caption{Optimum values of coherence parameters as a function of (a) the mechanical occupation number $n_m$ when $P_{ee} = 0$ and (b) the qubit probability amplitude $ P_{ee}$ when $n_{m} = 0.5 $, in the presence and in the absence of the coupling term $g_m$. Inset plots in panels (a) and (b) show the evolution of the Bloch vector during $ \omega_m t \in [0,30] $ for (a) different mechanical occupation numbers $n_m = 0$ and $n_m = 15$ with $P_{ee} = 0$, as well as (b) different values of $ P_{ee} =0$ and $ P_{ee}=0.48$ with $n_m = 0.5$. The evolution of $\mathcal{C}_q$ for (c) $P_{ee} = 0$ and (d) $P_{ee} = 0.48$ together with the quadratic fittings in the absence and in the presence of $g_m$, when $n_m = 0.5$. Inset plots in panels (c) and (d) show the zoomed rectangle-region of fitting for a short time interval $\omega_m t \in [0,1]$. (e) The evolution of $\mathcal{C}_m$ for two different values of $P_{ee} = 0$ and $P_{ee} = 0.48$ with and without coupling constant $g_m$. Other numerical parameters are the same as those in Fig.~\ref{fig2}.}
	\label{figA2}
\end{figure}
\section{Effects of the coherent driving term on the quantum coherence}\label{A2}
In the vicinity of the degeneracy point ($n_g \rightarrow 1/2$), the coupling term $g_m = g_0 (1- 2 n_g)$ is too small in comparison with other coupling rates $g_x$ and $g_z$ so that, the QID term can slightly modify the dynamics of the quantum coherence.
By considering the shift term $ g_m X_m $, the interaction Hamiltonian of the system now becomes
\begin{equation}\label{Hint_shift}
H_{int} = \left( g_x  \sigma_x + g_z \sigma_z + g_m \right) X_m.
\end{equation}
In Fig.~\ref{figA2}(a,b) the optimum values of the coherence parameters $ \mathcal{C}_q $ and $ \mathcal{C}_m $ with respect to the mechanical-thermal number $n_m$ (Fig.~\ref{figA2}(a)) and the qubit weight  $P_{ee}$ (Fig.~\ref{figA2}(b)) are depicted in the presence and the absence of the coupling rate $g_m$. From those panels, we can see that the constant shift moderately improves the results for both qubit coherence and mechanical coherent displacement.
In addition, the inset plots in panels (a) and (b) of Fig~\ref{figA2} show the evolution of the Bloch vector for different values of $n_m$ (Fig.~\ref{figA2}(a) ) and $P_{ee}$ (Fig.~\ref{figA2}(c)) when we consider the coupling term $g_m \neq 0$. As is evident from inset plots of Fig.~\ref{figA2}(a), by increasing the mechanical occupation numbers $n_m$, the expectations $ \langle \sigma_x (t) \rangle $ and $ \langle \sigma_y (t) \rangle $ take larger values which give rise to larger amount of qubit coherence parameter $\mathcal{C}_q $. On the other hand, increasing the qubit thermal number $n_q$ or equivalently, $P_{ee}$, causes the reduction in mean values of  $ \langle \sigma_x (t) \rangle $ and $ \langle \sigma_y (t) \rangle $ and consequently $\mathcal{C}_q $ (see inset plots in Fig~\ref{figA2}(b)). These results are completely in agreement with the previous outcomes explained in the body of the manuscript.
Moreover, in Panels (c,d) and (e) of Fig.~\ref{figA2}, the evolution of $\mathcal{C}_q $ and $\mathcal{C}_m $ in the absence and the presence of the QID term are depicted as a function of normalized time $ \omega_m t $ for two different values $P_{ee} = 0$ and $P_{ee} = 0.48$.
In accordance with Fig.~\ref{figA2}(a,b), the presence of $g_m$ can slightly change the values of the coherence parameters in time.

Similar to what we get in Eq.~(\ref{xpsxy}), we can also calculate the coherence components  for a very tiny time interval, when we apply QID $g_m X_m $ into the dynamics of the system
\begin{subequations}\label{xpsxy_A2}
\begin{align}
\langle {X}_m (t) \rangle &  \approx  \sqrt{2} t^2
  \left[ g_z \left( \omega_m ( 2 P_{ee} - 1 ) n_m ( 4 n_m + 3 ) + \omega_q ( n_m + \tfrac 12 ) \right)
  +
  g_m \left( \omega_m n_m ( 4n_m + 3 ) + \omega_q ( 2 P_{ee} - 1 ) ( n_m + \tfrac 12 ) \right)
  \right],
  \\
  \langle {P}_m (t) \rangle &  \approx  - \sqrt{2} g_z t (2 P_{ee} - 1), \\
  \langle {\sigma}_x (t) \rangle &  \approx  2 g_x t^2 (2 n_m +1) \left[ g_z \left(2 P_{ee} - 1 \right) + g_m \right], \label{xpsxy_A2_c}\\
  \langle {\sigma}_y (t) \rangle &  \approx  0.
\end{align}
\end{subequations}
As is evident from~\cref{xpsxy_c,xpsxy_A2_c}, the qubit coherence parameter evolves quadratically in a very short time domain. By introducing the fitting function $ F_{\mathrm{fit}}^{(g_m =0)} =| 2 g_x g_z (2 n_m + 1 )( 2 P_{ee} -1) t^2 | $ and $ F_{\mathrm{fit}}^{(+g_m)} =| 2 g_x (2 n_m + 1 ) (g_z ( 2 P_{ee} -1) + g_m) t^2 | $ associated with the  Eqn.~(\ref{xpsxy_c}) and (\ref{xpsxy_A2_c}), in Panels (c) and (d) of Fig.\ref{figA2} and their insets we checked the consistency of the analytical and numerical results for qubit coherence in two conditions of the absence and the presence of the coupling rate $g_m$, respectively. As can be seen, for a short time interval the results are matched which confirm that qubit coherence parameter behaves quadratically for the initial time interval.

\bibliography{coherences}
\end{document}